\documentclass{eas}
\usepackage{graphicx}
\newcommand{\msun}{\,\mbox{$\mbox{M}_{\odot}$}}

\TitreGlobal{The Ages of Stars}
\begin{document}

%%-----------------------------
%%      the top matter
%%-----------------------------
\title{Using rotation, magnetic activity and lithium to estimate the ages of low
mass stars} 
\author{R. D. Jeffries}\address{Astrophysics Group, Keele University,
  Keele, Staffordshire, ST5 5BG, UK \\ r.d.jeffries@keele.ac.uk}

\runningtitle{Jeffries: Rotation, activity and lithum}
%\author{...}\address{...}
%\author{...}\address{...}
%
%
\begin{abstract}
The rotation rate, level of magnetic activity and surface lithium
abundance are age-dependent quantities in stars of
about a solar mass and below. The physical reasons for the evolution of
these phenomena are qualitatively understood, but accurate quantitative
models remain dependent on empirical calibration using the Sun and
stars of known age, chiefly in clusters. In this work I review the
status of these ``empirical age indicators'', outlining the
astrophysics of their time dependence, describing the measurements,
assessing the precision (and accuracy) of age estimates when 
applied to individual stars, and
identifying their principle limitations in terms of the mass and age
ranges over which they are useful. Finally, I discuss the ``lithium
depletion boundary'' technique which, in contrast to the empirical
methods, appears to provide robust, almost model-independent ages that
are both precise and accurate, but which is only applicable
to coeval groups of stars.
\end{abstract}
\maketitle
%%-----------------------------
%%      your text
%%-----------------------------
\section{Introduction}

\label{sec1}

The age of a star is, along with its mass and composition, the
most important quantity to know for testing ideas
concerning the evolution of stars, stellar systems (clusters and
galaxies) and also, by association, their circumstellar material and
exoplanetary systems. However, unlike mass and composition, we have no
direct means of measuring the age of any star but the Sun. 
The ages of other stars are inferred or estimated using a hierarchy
of techniques, which can be described as (see Soderblom 2010;
Soderblom \etal\ 2013) semi-fundamental, model-dependent, empirical or
statistical. 

Semi-fundamental techniques rely on
age-dependent phenomena where the physics is understood, there is
little tuning of model parameters required and the results are
basically model-independent. Model-dependent techniques 
include isochrone fitting in the Hertzsprung-Russell (HR) diagram,
asteroseismology and white dwarf cooling. Here the physics is mostly
understood, but there are annoying gaps in our ability to accurately
model the physics without making simplifying assumptions or tuning
parameters (e.g. the mixing length) to match observations. Often the
precision of the ages determined by such techniques is much better than
their absolute accuracy and different models may yield ages that
differ by more than their claimed uncertainties.

At a level below the model-dependent techniques are
empirical age indicators. For these, the understanding of the physics
is qualitative, with significant holes in the theory that are usually
bridged using semi-empirical relationships with 
free parameters. The general approach is to calibrate an
age-dependent phenomena using similar observations of stars with
``known'' age (the Sun and stars with ages estimated by
semi-fundamental or model-dependent techniques) and then use that
calibration to estimate the ages of other stars (e.g. Barnes 2007;
Vican 2012). Of course, there is a
risk of circularity here; one cannot study the age dependence of a
phenomenon using stars with ages estimated using that phenomenon!

%Finally there are statistical techniques that can be used to infer an
%age, such as those based on the kinematics or abundances of stars. These
%are based on our understanding of galaxy evolution, rely on
%calibrations established using techniques above them in the hierarchy
%and are rarely of useful precision when applied to individual stars rather
%than populations.

In this contribution I deal mainly with empirical age indicators
associated with the rotation rates, levels of magnetic activity and
photospheric lithium abundances of stars with masses $M\leq 1.3$\msun\
and how they apply to stars from birth to ages of 10\,Gyr.
It is no coincidence that these phenomena each become useful below this
mass. The presence of a sub-photospheric convection zone is responsible
for dynamo-generated magnetic fields that are dissipated to provide
non-radiative heating in the outer atmosphere and also couple to an
ionised wind that drives angular momentum loss. The same convection
zone is responsible for mixing pristine material down to interior
regions where Li can be burned. The use of these indicators has its root
in work done by Kraft and collaborators in the 1960s (e.g. Kraft \&
Wilson 1965; Kraft 1967), but perhaps the
most influential early paper was by Skumanich (1972), who showed that both
rotation and activity, and to some extent Li abundance, decayed
according to the inverse square root of age. The data used were sparse,
consisting of the Sun (age 4.57\,Gyr) 
and a few solar-type stars in the
Pleiades (age $\simeq 125$\,Myr) and Hyades (age $\simeq 600$\,Myr)
open clusters, but nevertheless this paper stimulated much of what
follows.

The utility of these empirical age indicators is mostly in estimating
ages for low-mass main sequence (MS) and pre main sequence (PMS)
stars that constitute the vast 
majority of the Galactic population. A principle advantage of the
techniques I will discuss is that they are {\it distance
  independent}. With the successful launch of the {\it Gaia} satellite
(Perryman \etal\ 2001; Brown 2008), it might seem that uncertain stellar
distance will be a solved problem within a few
years. However, even with precisely known distances, the determination
of ages for stars that have reached the main sequence and are still
burning hydrogen in their cores is difficult. Position in the 
HR diagram is age sensitive, but also sensitive to the
detailed composition of the star. Even with [Fe/H] known to a
very respectable accuracy of $\pm 0.05$\,dex, the age of a 5\,Gyr
solar-mass star could only be estimated to a precision of 20 per cent,
and considerably worse for lower mass stars with longer main sequence
lifetimes that consequently move more slowly in the HR diagram (e.g. see Fig.~20 of Epstein \& Pinsonneault
2014). Asteroseismology may be an alternative distance-independent
method for age estimation, with the advantage of a strong and
well-understood physical basis, but it is not clear that pulsations
can easily be detected in main-sequence stars well below a solar mass
or in young,
active stars (e.g. Huber \etal\ 2011). Even if they are,
it is unlikely that ages could presently be estimated for solar-type
stars to absolute
precisions better than
10--15 per cent of their main sequence lifetimes (e.g. Gai \etal\ 2011;
Chaplin \etal\ 2014) and would rapidly become too large to be
useful in stars below a solar mass. 
Hence, there is likely to be a
need for age determinations using empirical indicators for the
forseeable future.

In section~\ref{sec2} I discuss measurements of rotation in low-mass
stars, the physical basis on which rotation rate could be used to
estimate age and review efforts to calibrate ``gyrochronology''. 
Section~\ref{sec3} reviews the connection between rotation and magnetic 
activity and the various attempts to calibrate activity-age
relationships using several magnetic activity
indicators. Section~\ref{sec4} discusses the astrophysics of lithium
depletion in solar-type stars, comparison of observations and models
and the use of lithium as an empirical age indicator in PMS and MS
stars separately. Also included is a description of the ``lithium
depletion boundary''
technique in very low mass stars, which differs from the other methods discussed here in that
it requires no empirical calibration and is semi-fundamental. 
Section~\ref{sec5} summarises the status and
range of applicability of each of these techniques and briefly discusses efforts
to improve empirical calibrations. Conclusions are presented in section~\ref{sec6}.

\section{Rotation rates and gyrochronology}

The motivation for using rotation rate as an empirical age indicator is
discussed extensively by Barnes (2007). As well as being
distance-independent it seems, at least for older stars (see below), 
there may be an
almost unique relationship between rotation rate and age. Rotation
rates are increasingly available; satellites
such as  {\it CoRoT} and {\it Kepler} have accumulated large quantities
of rotation data (Affer \etal\ 2012; McQuillan, Mazeh\& Aigrain 2014), and ground-based experiments such as {\it HATNet} and {\it SuperWASP}, aimed primarily at
variability or 
exoplanet searches, have the potential
to provide rotation periods for vast numbers of stars
(e.g. Hartman \etal\ 2011; Delorme \etal\ 2011). Photometric
monitoring by {\it Gaia} will add to this haul.

\label{sec2}

\subsection{Measuring rotation rates}

Rotation rates in low-mass stars can be found in a
number of ways (see the review by Bouvier 2013), but only two are
mentioned here; the others are generally more difficult to
apply routinely. Spectroscopy can be
used to estimate that component of spectral line broadening
contributed by rotation -- the projected equatorial velocity, $v sin
i$. This can be accomplished by a direct Fourier transform of the
spectrum (e.g. Gray 1976; Dravins, Lindegren \& Torkelsson 1990) and 
with very high quality data, it can even be feasible to measure differential
rotation with latitude (e.g. Reiners 2007). More frequently, $v \sin i$
is estimated by calibrating the width of a cross-correlation peak against
template stars or synthetic spectra with similar atmospheric parameters
(e.g. Rhode, Herbst \& Mathieu 2001). 
Although feasible using a single spectrum, 
and in the case of
cross-correlation, a spectrum with very modest signal-to-noise
ratio, the principle limitations of spectroscopic methods are the 
high resolving powers
and accurate characterisation of the intrinsic (non-rotating) line
profiles required to estimate $v\sin i$ for slow rotators, and the
confusing $\sin i$ axis orientation term. 

The main alternative, and method of choice, is to monitor the brightness of stars and detect
periodic modulation caused by dark magnetic starspots or bright
chromospheric plages on their surfaces. 
Magnetic activity is required for this technique to work, so is best
suited to low-mass stars at younger ages with vigorous rotational
dynamos (see section~\ref{sec3}), where typical modulation amplitudes 
can be a few
mmag to tenths of a magnitude. Typical examples of such studies can be
found in Prosser \etal\ (1993), 
Allain \etal\ (1996), Herbst \etal\ (2000) and   
Irwin \etal\ (2009), which also demonstrate a progression in the efficiency
of monitoring facilitated by the advent of large format CCDs. The
principle advantage of this technique is that many stars
can be almost simultaneously 
monitored using telescopes of modest aperture (compared with
those required for spectroscopy). The disadvantages are that stars need
to be monitored intensively and over at least a couple of rotation
periods. There is also a potential bias towards the young, most active and
most rapidly rotating stars -- even in young, magnetically active
cluster populations $\leq 50$ per cent of stars have measured
rotation periods and older stars may have such small photometric
amplitudes and long periods that only space-based photometry is good enough. Both the {\it Kepler} and {\it CoRoT} satellites have
provided much more precise, lengthy time-series data for 
field stars (and some clusters) to partly nullify these
problems (e.g. Meibom \etal\ 2011a; Affer \etal\ 2012; Reinhold,
Reiners \& Basri 2013; McQuillan \etal\ 2014).

Prior to {\it Kepler} and {\it CoRoT}, most information about the rotation of
older, less active stars came from chromospheric inhomogeneities
monitored in the Ca\,{\sc ii} H and K lines (e.g. Donahue, Saar \&
Baliunas 1996; Baliunas \etal\ 1996), because the
contrast between chromospheric plages and the immaculate photosphere is
greater than for starspots in stars as old/inactive as the Sun.
Monitoring on decadal timescales at the Mount Wilson
observatory has yielded many rotation periods for solar-type field
stars as well as quantitative measurements of their magnetic activity
and magnetic activity cycles (see section~\ref{sec3}). 

\subsection{Rotational evolution and models} 

\begin{figure}

\begin{center}
\includegraphics[width=10cm]{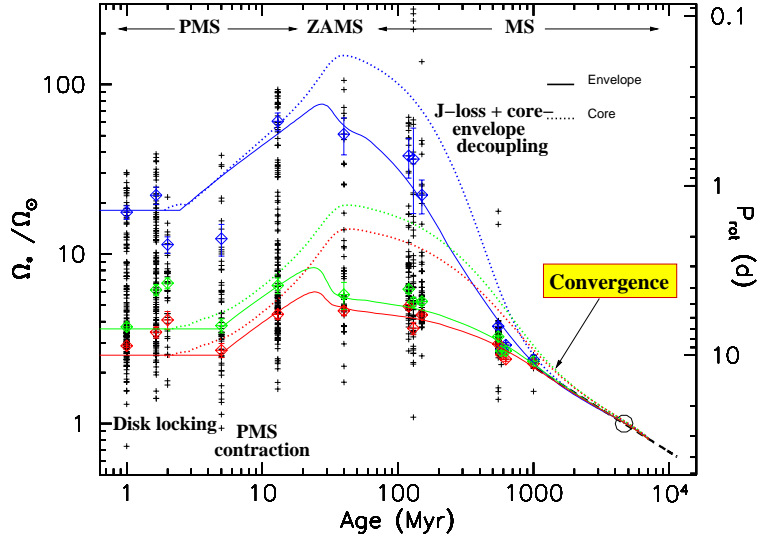}
\end{center}
\caption{Rotation rates/periods for sets of solar-type stars in coeval
  clusters as a function of age (adapted from Gallet \& Bouvier 2013).
The PMS, ZAMS and MS phases are marked and the dominant physical
processes at work are indicated. Beyond ages of $\sim 0.5$\,Gyr rotation
rates converge for
stars of a solar mass, or at least are predicted to converge, to a
close-to-unique function of age. This convergence takes longer at lower
masses.}
\label{gallet13}
\end{figure}

\subsubsection{Observed rotational evolution}

Most progress in understanding the rotational evolution of solar-type
and lower-mass stars comes from observations of rotation rates
(predominantly rotation periods) in clusters of stars, whose members
are assumed coeval and of similar composition. Compilations of
data and reviews of the observations can be found in Irwin \& Bouvier
(2009), Gallet \& Bouvier (2013) and Bouvier \etal\ (2013), and these sources
also provide an overview of theoretical
interpretations of these observations.
Figure~\ref{gallet13} (from Gallet \& Bouvier 2013) illustrates the
main features of rotational evolution for groups of stars at around a
solar mass, ranging in
age from star forming regions at a few Myr, through to the ZAMS at
$\sim$ 100\,Myr and onto later main sequence life beyond a Gyr. 

Solar-type stars
evidently begin their lives with a wide range of rotation periods
between about 1 and 15 days (e.g. in the Orion Nebula cluster; Herbst
\etal\ 2002, or NGC~2264; Makidon \etal\ 2004). Over the first 10\,Myr of their lives this distribution
changes little despite the order of
magnitude reduction in moment of inertia as stars contract along
their PMS tracks. Interactions between the star and its circumstellar
disk are invoked to remove angular momentum, a process that
ceases upon the dispersal of inner disks on timescales of a few
Myr. This idea finds support
from the correlation found in some star forming regions between
the presence of disks/accretion and slower rotation (e.g. Edwards
\etal\ 1993; Rebull \etal\ 2006; Cieza \& Baliber 2007). 

The rotation rate distributions in older clusters
show gradual evolution towards faster rotation rates at the ZAMS,
presumably as a
result of PMS contraction. Although the long-period envelope remains 
fairly constant in solar-type stars, there are few slow rotators among
lower mass ($\leq 0.5\,M_{\odot}$) stars at ages of 10-200\,Myr, with
most having rotation period $P<3$\,d (e.g. Irwin \etal\ 2007, 2008). The range of rotation
rates in solar-type stars 
rapidly increases to nearly two orders of magnitude; at $\sim 15$\,Myr
the rotation rate distribution is still quite flat but the range has
grown to $0.2<P<15$\,days (e.g in the h Per cluster; Moraux
\etal\ 2013). At $\sim 50-150$~Myr the bulk
of solar-type ZAMS stars have $6<P<10$\,days, but a tail of rapid rotators
persists to periods as short as 0.3 days (e.g. in the Alpha Per and Pleiades
clusters; Prosser \etal\ 1993; Krishnamurthi \etal\ 1998; Hartman
\etal\ 2010). 

Beyond the ZAMS, with the moment of intertia essentially
fixed, the wide distribution of rotation rates
in solar-type stars converges, a process thought to be driven by a magnetised
stellar wind, with angular momentum losses that increase with rotation
rate. Convergence is almost complete for solar-type
stars at ages of $\geq 500$\,Myr (e.g. in the Hyades; Radick
\etal\ 1987, or in M37; Hartman \etal\ 2009). The timescale for
convergence is however mass-dependent and fast rotating K-dwarfs are
still seen in clusters with ages of a few hundred Myr (e.g. in M34; Meibom
\etal\ 2011b), whilst M-dwarfs with
rotation periods $<1$\,d are still observed in the Hyades and Praesepe
clusters at ages of $\sim 600$\,Myr (Delorme \etal\ 2011; Ag\"ueros
\etal\ 2011). In fact if anything, the dispersion in rotation rates
appears to grow with age in these lower-mass stars as evidenced in the
wide range of periods found for (predominantly old) field M
dwarfs (Irwin \etal\ 2011; McQuillan, Aigrain \& Mazeh 2013). 

\subsubsection{Rotational evolution models}

Models to interpret these data are semi-empirical; there
are several components that, whilst physically motivated, require
calibration using cluster data and the current rotation rate of the
Sun. Starting from an initial rotation period
at a very young age, the effect of torques and moment of inertia
changes are followed and models include some or all of the following
ingredients:

\noindent
{\bf Star-disk interactions:} There is no general
  agreement yet on which mechanisms prevent the spin-up of
  contracting PMS stars, but the presence of an inner
  disk appears to be implicated. The necessary transfer of
angular momentum may be provided via the original ``disk-locking'' proposed 
between the accretion disk and stellar magnetic field (Camenzind 1990; 
Koenigl 1991); more recent ideas include 
accretion-driven winds or magnetospheric
ejections (e.g. Matt \& Pudritz 2005; Zanni \& Ferreira 2013). 
Whatever is responsible, most rotational evolution models assume that
  rotation rates are held constant until the inner
  disk disperses. This disk dispersal timescale, observationally found
  to be in the range 1--10\,Myr, almost certainly
  varies from star-to-star for poorly understood reasons and is a tuneable
  model parameter, largely set by the difference in the mean and range
  of rotation rates at the ZAMS compared with those in the 
  initial distribution (e.g. Bouvier, Forestini \& Allain 1997).

\noindent
{\bf PMS Contraction:} Once disks disperse then stars are 
free to spin-up if they have not reached the ZAMS. 
The moment of inertia will decrease roughly on
the Kelvin-Helmholtz timescale -- around 10\,Myr for a solar
mass star, but hundreds of Myr for lower-mass stars (i.e.
much longer than any disk dispersal timescale). Stellar (surface) spin up 
is moderated both by
angular momentum losses and the possible decoupling of radiative core
and convective envelope (see below).

\noindent
{\bf Wind angular momentum loss:} 
  The large scale magnetic B-field of a
  star will force its ionised stellar wind into co-rotation out to
  some distance approximated by the Alfven radius. Upon decoupling, the
  wind carries away angular momentum at a rate that depends on the
  rotation rate of the star, the mass-loss rate, the strength and
  geometry of the magnetic field and the details of the wind velocity
  profile and interaction with the magnetic field (Mestel \& Spruit
  1987). A common parametrisation attributable to Kawaler (1988) and
  Chaboyer, Demarque \& Pinsonneault (1995a) is
\begin{equation}
\frac{dJ}{dt} = f_k K_w \left(\frac{R}{R_{\odot}}\right)^{2-N}
  \left(\frac{M}{M_{\odot}}\right)^{-N/3}
    \left(\frac{\dot{M}}{10^{-14}}\right)^{1-2N/3}
    \Omega^{1+4N/3}\ \ (\Omega < \Omega_{\rm crit})\, , \\
\label{jdot}
\end{equation}
\begin{equation}
\frac{dJ}{dt} = f_k K_w \left(\frac{R}{R_{\odot}}\right)^{2-N}
  \left(\frac{M}{M_{\odot}}\right)^{-N/3}
    \left(\frac{\dot{M}}{10^{-14}}\right)^{1-2N/3}
    \Omega\, \Omega_{\rm crit}^{4N/3}\ \ (\Omega \geq \Omega_{\rm crit})\, , 
\label{jdotsat}
\end{equation}
where $N$ is an index specifying the B-field geometry ($N=2$ is
radial, $N=3/7$ represents a dipolar field), $\dot{M}$ is the
wind mass-loss rate in solar masses per year, $K_w$ is a
constant ($=2.036\times10^{33}$ in cgs units), $f_k$ is a parameter
that encapsulates the constant of proportionality in an assumed linear 
relationship
between surface magnetic {\it flux} and rotation rate $\Omega$, 
as well as uncertainties in the wind
speed as it decouples from the field at the Alfven radius. The strong
dependence on $\Omega$ is the main physics behind the convergence of
rotation rates in later main sequence life. 
$\Omega_{\rm
  crit}$ is a threshold rotation rate at which the B-field and
consequently the angular momentum loss rate ``saturate''. This is
motivated by the need to ensure that fast-rotating stars do not
spin-down too quickly upon reaching their ZAMS radius and the
observation that saturation
is observed in chromospheric and coronal indicators of magnetic
activity (see section~\ref{sec3}).

Of these parameters, several need to be assumed (e.g. $N$, the
relationship between B-field and rotation rate) or fixed by ensuring
that at 4.5\,Gyr, the rotation rate of the Sun is reproduced (e.g. $f_k$). If
$N\simeq 1$ the angular momentum loss rate is not too dependent on the assumed
$\dot{M}$, which is fortunate as there are few constraints on this
for stars younger than the Sun. Some of these degrees of freedom are
beginning to be constrained by new MHD simulations,
albeit still with simplifying assumptions about B-field geometry
(e.g. Matt \etal\ 2012). Using equation~\ref{jdot} for a star
with a fixed moment of inertia and $\Omega < \Omega_{\rm crit}$ 
leads directly to $\Omega \propto
t^{-\alpha}$, with $\alpha=1/2$, as suggested by Skumanich (1972). 
However, it is of
critical importance in what follows to note that the $t^{-1/2}$
behaviour is very dependent on model assumptions and is by no means
assured of applying at all masses. For instance Reiners \&
Mohanty (2012) have pointed out that if there is instead a linear relationship
between magnetic {\it field} and rotation rate then the radius
dependence of $dJ/dt$ is much stronger (e.g. $\propto R^{16/3}$ for a
radial field rather than radius-independent in equation~\ref{jdot}). As
$R$ changes even during main sequence evolution, this changes the form
of $\Omega(t)$. Similarly, any mass-dependent or time-dependent changes
in $\dot{M}$ or B-field topology will alter $\alpha$ and
possibly give it a mass- or time-dependence.

\noindent
{\bf Core-envelope decoupling:}
As angular momentum is lost from the surface, interior processes
act to transport angular momentum within stellar radiative zones - these may include
hydrodynamic instabilities, magnetic fields or gravity waves (Mestel \&
Weiss 1987; Chaboyer \etal\ 1995a; Pinsonneault 1997; Mathis
\etal\ 2013; Mathis 2013; Charbonnel \etal\ 2013). Some
studies treat this numerically as a diffusion process within radiative zones
Denissenkov \etal\ 2010; Eggenberger \etal\ 2012), others allow the radiative core
and convective envelope to rotate as solid bodies at different rates
with a coupling timescale (e.g. MacGregor \& Brenner 1991; 
Gallet \& Bouvier 2013). In either case
there are free parameters associated with the diffusion coefficients or coupling
timescales that can partly be constrained by what we know about the
internal rotation of the Sun and also its surface lithium abundance
(see section~\ref{sec4}),
but otherwise must be considered free parameters that may depend
on mass or surface rotation rate.

\subsubsection{Putting it together}
Parametrised models incorporating some or all of these features have been
studied by many authors in the past decades; more recent studies include 
Denissenkov \etal\ (2010), Spada \etal\ (2011) and Gallet \& Bouvier
(2013). Models begin with an assumed rotation rate for young stars and
typically assume
this rotation rate is constant whilst the star possesses an inner
disk. A prescription for wind angular momentum losses and a stellar
evolution model are then used to follow rotation rate as the star
contracts and loses angular momentum. The core and envelope may be
treated as separate solid body rotators with a coupling timescale.
There are sufficient free or assumed parameters in this model (e.g. the dynamo
prescription, initial rotation rate, disk lifetime, coupling timescale,
$f_k$, $\Omega_{\rm crit}$) that reasonable fits can be found to the
observed distribution of rotation rates. Another degree of freedom is
the mass-dependence of these parameters. It has been known for some time
that models which match the evolution of rotation rate distributions
for solar-type stars do not adequately match those of lower mass stars
using the same set of parameters (e.g. Sills, Pinsonneault \& Terndrup 
2000; Irwin \etal\ 2011). A
mass-dependent $\Omega_{\rm crit}$, changes in magnetic topology or 
dynamo location, or
wind braking laws with a more
extreme mass/radius dependence (e.g. Reiners \& Mohanty 2012; Brown 2014) may
provide solutions, but at the moment we are some way from being able to
directly estimate stellar ages from models of rotational evolution.

\subsection{Empirical gyrochronology}

\subsubsection{Gyrochrone construction and use}
\label{gyrouse}

\begin{figure}
\begin{center}
\includegraphics[width=10cm]{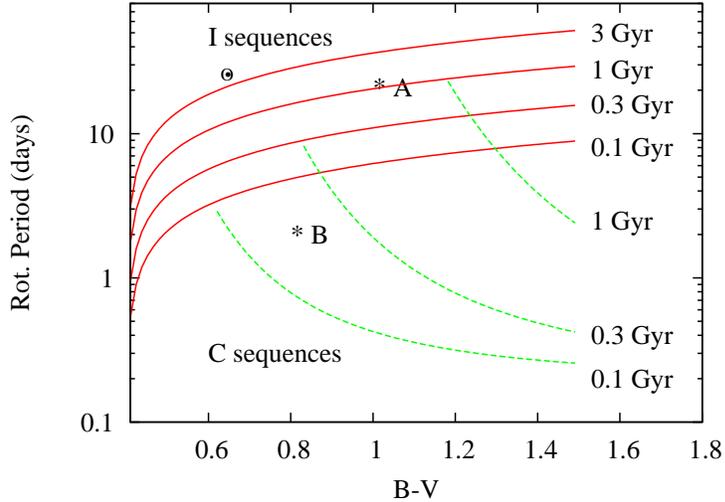}
\end{center}
\caption{A schematic of the location of the I- and C-sequences in the
  rotation period, colour plane. The I-sequences (gyrochrones) were
  calculated according to the formula advocated by Barnes (2007), the
  C-sequences from the formula given by Barnes (2003). The Sun and two
  illustrative stars (A and B) are shown and discussed in the text.}
\label{gyrochrone}
\end{figure}

The observed evolution of rotation rate distributions and a
semi-empirical understanding of the processes involved offer both
problems and opportunities. The broad range of rotation rates seen in
solar-type stars between about 1 and 200\,Myr, and to even older ages for lower
mass stars, mean that rotation rate is a {\it poor} age indicator for
individual stars at these ages. However, the model predictions of a
convergence in rotation rates to a single-valued function of age for
older main sequence stars, and observations that suggest this may
actually happen at least to ages of a Gyr or so in solar-type stars,
suggest that rotation {\it can} be a good empirical age 
indicator for these stars.

Barnes (2003, 2007) noted that when plotted in
the $P$ versus $B-V$ plane, stars in young ($<1$\,Gyr) clusters
and older field stars appear to populate two distinct
sequences (schematically illustrated in Fig.~\ref{gyrochrone}) and which
Barnes termed the I- and C-sequences. The I-sequence appears
well-established in a number of clusters and is formed from those stars
with converged rotation rates, roughly following the $\Omega \propto t^{-1/2}$
Skumanich law. The C-sequence appears less tight and is populated
by rapidly rotating stars, which are in the saturated regime of
magnetic activity (see section~\ref{rot_activity}) with $\Omega >
\Omega_{\rm crit}$. As the convergence timescales are observed to be
longer in low-mass stars, the junction of these two sequences moves
redward, with everything blueward of the junction being on the I-sequence. 
Barnes suggested that stars on the I-sequence have a fully (magnetically)
coupled core and envelope, whereas in stars on the C-sequence the core
and envelope are decoupled. The dearth of objects between the two sequences
reflects a short evolutionary timescale between the two states and
Barnes (2003) interpreted this as a switching between a convective (C) and an
interface (I) dynamo that couples the core and envelope, causing a
rapid spindown.

Irrespective of the (strongly debated -- e.g. Denissenkov \etal\ 2010; Barnes \&
Kim 2010; Brown 2014) physical processes at play, the phenomenon of an
I-sequence can be used to estimate stellar ages. The basic procedure is
to say that rotation rate is a {\it separable} function of both age and
colour/mass/$T_{\rm eff}$. i.e.
\begin{equation}
P(B-V,t) = f(B-V)\,g(t)\, ,
\label{gyroeq1}
\end{equation}
\begin{equation}
f(B-V) = a\,[(B-V) - b\,]^c\, , \ \ {\rm and} \ \ \ g(t) = t^\alpha\, .
\label{gyroeq2}
\end{equation}
The function $f$ is chosen to match the shape of the I-sequence in
young clusters, whilst $\alpha=0.5$ is equivalent to the Skumanich
spin-down law. A number of authors have calibrated relationships of
this type (e.g. see Table~1 in Epstein \& Pinsonneault 2014); $f$ is found
from fitting one (or several) clusters in the $P$ vs $B-V$ plane,
whilst $\alpha$ is determined by matching the solar rotation
rate. Different authors find $\alpha$ in the range 0.52--0.57 
and there are significant differences in the form of $f$ too (Barnes
2007; Mamajek \& Hillenbrand 2008; Meibom, Mathieu \& Stassun 2009, 2011a; Collier
Cameron \etal\ 2009).

Figure~\ref{gyrochrone} shows gyrochrones (I-sequences) calculated from
equations~\ref{gyroeq1} and~\ref{gyroeq2}, with the parameters derived
by Barnes (2007), and C-sequences generated according to the functional 
form defined by Barnes (2003). Two hypothetical stars, A and B,
are shown along with the Sun. Star A lies just above the 1\,Gyr
I-sequence gyrochrone {\it and} to the left of the C-/I-sequence junction for
1\,Gyr. Therefore its age can be estimated as just greater than
1\,Gyr. Star B however lies well below the 0.1\,Gyr I-sequence but {\it
  to the
right} of the 0.1\,Gyr C-sequence. As stars with ages up to about 0.3\,Gyr could
exist at this period/colour (on or above their C-sequences) then about
the best we can say as that star B is $<0.3$\,Gyr. This is a fundamental
limit of gyrochronology that traces back to the large scatter in
rotation rates seen at ages prior to the (mass-dependent) convergence.

\subsubsection{Problems, precision and accuracy}

\label{gyrolimitations}

In addition to the fundamental problem at young ages just discussed, it
can be difficult to measure the rotation period of a star. Whilst young
stars (or those with short periods) {\it may} have a healthy photometric modulation amplitude that
can be detected from the ground (actually censuses of rotation period
are often $<50$ per cent complete even in young clusters), 
this is not generally true at older
ages, where precision photometry from space is required (see Fig.9
in~Reinhold, Reiners \& Basri 2013). Alternatively,
chromospheric modulation can be used in older stars (e.g. Donahue
\etal\ 1996), but is {\it much}
more expensive in terms of observing time.
The precision and accuracy of gyrochronology may also be affected by measurement
uncertainties, differential rotation, limited convergence, binarity and most
importantly, calibration uncertainties.

\noindent
{\bf Differential rotation:} The precision of most period
measurements is high, but stars may have differential rotation with
latitude such that the period measured at one epoch may not be that
measured at another, depending on the latitudinal starspot
distribution. Differential rotation has been studied using the Mt
Wilson chromospheric activity time-series, finding $\Delta P/P \simeq
0.05\, P^{0.3}$ (Donahue \etal\ 1996; where
$\Delta P$ is the range of periods found at a single epoch). Reinhold
\etal\ (2013) use Kepler data to show that, compared with the Sun (which
has $\Delta P/P \sim 0.14$) differential rotation increases with
$T_{\rm eff}$ and $P$. If $\Delta P/P=0.1$ and $P \propto t^{-1/2}$,
then this leads to an age uncertainty of 20 per cent if only a
single measurement of $P$ is available.

\noindent
{\bf Limited convergence:} The assumption of convergence to a unique
I-sequence is approximate. Convergence takes longer at lower masses and so
the older and more massive the star, 
the better this approximation is. Epstein \& Pinsonneault (2014)
perform simulations and show that this dominates over likely
uncertainties caused by differential rotation at $M<0.7\,M_{\odot}$ and
grows very rapidly below $0.4\,M_{\odot}$. Conveniently, convergence
becomes a problem in stars where differential rotation is unlikely to
be important and vice-versa, so the overall precision of gyrochronology
is likely abut 20 per cent in most cases. The situation may be a little
better in stars with ages of 0.5--1\,Gyr. The empirical
scatter of rotation periods around the I-sequence
in such clusters suggest that the combination of differential
rotation and incomplete convergence could lead to age errors of only 9--15
per cent (Collier Cameron \etal\ 2009; Delorme \etal\ 2011).

\noindent
{\bf Binarity:} Most of the discussion in this section applies to
single stars or at least stars that are effectively isolated from their
companions as far as angular momentum transfer is concerned. In
particular, stars in close, tidally locked binary systems may appear to
rotate at rates much faster than in a single star of similar
age.

\noindent
{\bf Calibration uncertainties:} Gyrochronology is essentially
calibrated using a group of young ($\leq 600$\,Myr) clusters and the
Sun. In particular, the assumptions of separable mass and time
dependencies and a simple, unique power-law time dependence may not be
true. For example different wind angular momentum loss prescriptions
give quite different predictions of the mass-dependence of $\Omega$,
even when tuned to match the rotation rate of the Sun (Reiners \&
Mohanty 2012). In addition, mass
or time-dependencies in $\dot{M}$, B-field topology and stellar radius
could lead to radically different gyrochrone shapes and spacing at
lower masses and older ages. Some confidence can be gained by noting that the
gyrochronological ages of the
components of a few (wide)
binary pairs with known rotation periods and differing masses
are roughly in agreement (Barnes 2007), but they have large individual
uncertainties and there are indications that these ages may not agree
with those from asteroseismology (Epstein \& Pinsonneault 2014).
There is an urgent need for better calibrating data (stars with
known age and rotation period) at lower masses than the Sun and at ages
of 1-10\,Gyr (see section~\ref{improvements}).

\section{Magnetic activity as an age indicator}

\label{sec3}

\subsection{Magnetic activity indicators}

Some of the difficulties associated with calibrating gyrochronology can
be addressed by using proxies for rotation that are easier to
measure. In stars with outer convection zones it appears that a
rotational dynamo can sustain a magnetic field, which emerges from the
photosphere and provides a source of non-radiative heating, leading to
the formation of a chromosphere and a hot corona.

Indicators of magnetic activity include coronal X-ray emission from gas
at $10^6$--$10^7$\,K. The chromosphere is at lower
temperatures but there are many emission lines that can be found which
act as diagnostics of the magnetic heating process(es), found mainly in
the blue and ultra-violet part of the spectrum, but which also include
Balmer-line emission. Each of these diagnostics demands different
technologies and techniques for their study and describing these is beyond the
scope of this review. Here we just need to know that usually, a {\it
  distance-independent} magnetic
activity index can be formed from the excess emission beyond that
expected from a normal photosphere, normalised by the bolometric
luminosity. The
principal examples in most of the literature on age determination are:
the Mt Wilson $R^{'}_{\rm HK}$ index, formed from the chromospheric flux
found in the cores of the Ca\,{\sc ii} H and K lines normalised by the
bolometric flux; and the ratio of X-ray to bolometric flux. 

Generally speaking, magnetic activity indices are easier to measure
than a rotation period and are often assessed with a single epoch of
observation -- though this can bring problems (see below). There are of
course limitations imposed by the sensitivity of instruments, the
distance to the stars in question and the contrast between the photosphere
and the magnetic activity indicator, which gets weaker as stars become
less active. i.e. Just like rotation, activity gets harder to measure
in older stars.

\subsection{The rotation-activity relationship}
\label{rot_activity}

The utility of magnetic activity as an age indicator arises because of
its close connection with rotation. This connection, ultimately due to the
nature of the interior dynamo that amplifies the magnetic field, has
been empirically understood for some time. For instance Pallavicini
\etal\ (1981) noted a good correlation between X-ray luminosity and the
square of the rotation velocity;  Noyes \etal\ (1984) found an
equivalent inverse correlation between flux in the Ca~{\sc ii}~H~and~K
lines and the rotation period. Noyes \etal\ also noted that a much
tighter correlation could be found between the ratio of chromospheric
flux to bolometric flux ($R^{'}_{\rm HK}$) and the inverse of the
Rossby number ($N_R$, the ratio of rotation period to turnover time at the
base of the convection zone). The turnover time increases with
decreasing mass and $N_R$ has become the parameter of choice in
activity-rotation correlations because it reduces the scatter when
combining data for stars with a range of masses and convective turnover times
(e.g. Dobson \& Radick 1989; Pizzolato \etal\ 2003; Jeffries \etal\ 2011; Wright \etal\ 2011).
 
At small $N_R$ (shorter rotation periods, or longer convective turnover times
at lower masses or in PMS stars) X-ray and chromospheric activity
indicators {\it saturate}. They reach a plateau at $N_R \leq
0.1$ below which they do not increase further, whilst at larger $N_R$,
activity decreases (Vilhu \& Walter 1987; Pizzolato \etal\ 2003). 
$N_R=0.1$ corresponds to $P
\simeq 3$\,d for a solar type star, but $P \simeq 6$\,d for an M0 star with a longer turnover
time. This saturation poses serious difficulties for the use of
activity as an age indicator in young stars. The period at which
saturation occurs in solar mass stars is
just below the I-sequence gyrochrone at 100\,Myr, so a large
fraction of stars at this age and younger have saturated magnetic
activity and therefore the observation of saturated magnetic activity
in a star can only yield an upper limit to its age. This age ambiguity
grows at lower masses because the increasing convective turnover times and
longer spin-down timescales of lower mass stars means that a larger
fraction of stars are saturated at a given age and they remain in the
saturated regime for longer.

\subsection{Empirical activity-age relationships}

\begin{figure}
\begin{center}
\includegraphics[width=10cm]{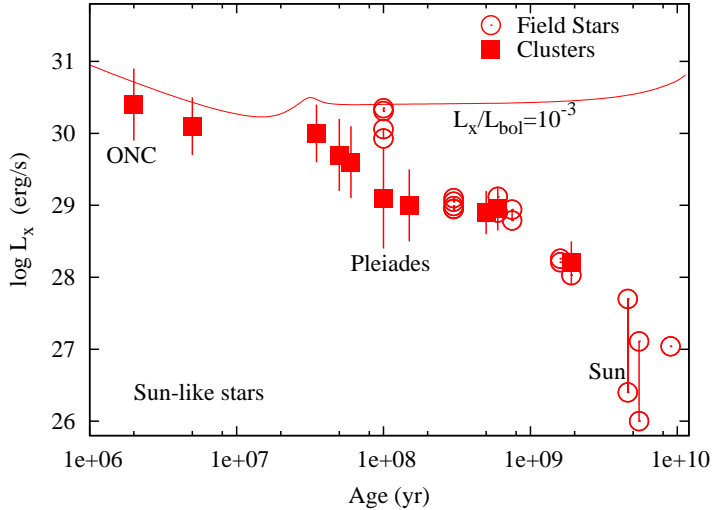}
\end{center}
\caption{The age-dependence of coronal X-ray luminosity for stars of
  about a solar mass. The median
  $L_{x}$ for clusters are shown with squares; the error bars indicate
  the interquartile range. Open circles are measurements of field stars
  with estimated ages. Lines connect observations of the same star at
  different epochs. The solid line indicates the locus where
  $L_{x}/L_{\rm bol}=10^{-3}$ (from the models of Siess, Dufour \&
  Forestini 2000), which is the observed saturation
  threshold for X-ray activity.}
\label{lxage}
\end{figure}

\subsubsection{X-ray activity}

Reviews of what is empirically known about the time-dependence of
magnetic activity can be found in Randich (2000), Ribas \etal\ (2005) and G\"udel (2007).
Figure~\ref{lxage} shows an empirical relationship between X-ray
luminosity and age for stars of about a solar mass. Different symbols
show mean levels and the interquartile range from surveys of open
clusters with ``known age'' 
(data are from Randich
2000; Flaccomio, Micela \& Sciortino 2003; Preibisch \& Feigelson 2005, and references therein) and data
for a few field stars and the Sun, where ages have been estimated by
other means, and where lines connect multiple measurements of the same
star (Telleschi \etal\ 2005; G\"udel 2007). These data are not complete, but they illustrate the basic
principles and problems of using empirical activity-age relations to
estimate ages. 

The overall decay of X-ray activity with age is clearly
seen, but the decay is not rapid for the first few hunded Myr,
especially if the X-ray luminosities were normalised by bolometric
luminosity to make them distance independent age indicators. In
addition there is scatter at all ages that cannot be attributed to
observational uncertainties. In the very young clusters there
is at least an order of magnitude range of $L_x$ (or $L_x/L_{\rm
  bol}$). This spread is not associated with rotation; most stars here
have very low Rossby numbers that would put them in the saturated
regime. Some of the spread may be associated with flaring or the presence of
circumstellar material (e.g. Wolk \etal\ 2005; Flaccomio, Micela \&
Sciortino 2012). In the young ZAMS clusters the spread in X-ray activity
remains, but this is known {\it not} to be due to variability
(e.g. Simon \& Patten 1998; Jeffries
\etal\ 2006) and these
stars have lost their circumstellar material . Instead, the rotation-activity correlation is at
work. Many stars have spun down below the threshold where their
activity is saturated and hence exhibit lower activity levels, whilst
other stars in the same cluster remain as rapid rotators. The lack of
variability and strong connection with rotation persists until at least
the Hyades at an age of 600\,Myr, with $L_{x}/L_{\rm bol}$ being
proportional to $N_R^{-2.7}$ in the unsaturated regime
(Wright \etal\ 2011).

Beyond 1\,Gyr we expect from Fig.~\ref{gallet13} that rotational
convergence has taken place, $\Omega \propto t^{-1/2}$, and hence
$L_{x}/L_{\rm bol} \propto t^{-1.35}$. If anything, the decay looks
like it may be a little steeper than this but there are no old open
clusters with good ages near enough to study in detail with X-ray
telescopes.  Ribas
\etal\ (2005) derived a time-dependence of between $t^{-1.27}$ and
$t^{-1.92}$ for the soft and hard X-ray fluxes from 6 solar analogues
in the field. Observations of field stars reveal, in contrast to the
younger stars, a high level of variability on timescales of years. The
Sun's soft X-ray emission changes by almost two orders of magnitude on
a roughly 11-year cycle (Strong \& Saba 2009) and there are now
observations of several solar analogues that indicate that at some point
beyond 1\,Gyr, large (order of magnitude) 
and possibly cyclic variability of X-ray emission
may commonly occur (e.g. Favata \etal\ 2008; Robrade \etal\ 2012).

In terms of mass dependence, the decay of coronal activity closely
follows what happens with rotation rates. Lower mass stars take longer
to spin down and remain in the saturated regime for longer. So, whilst
X-ray activity is a poor age indicator for the first 100\,Myr in a
solar-type star, this period extends to 1\,Gyr or more in M-dwarfs. The
longer spin down timescales also mean that field M-dwarfs tend to be
more active than field G-dwarfs (when expressed as $L_{x}/L_{\rm bol}$,
e.g. Preibisch \& Feigelson 2005). 

Even once stars have spun down and
reached the converged I-sequence of rotation periods, X-ray activity is
still not a very good age indicator because of the high levels of
variability. This limits precisions to about a factor of two in age.
In principle this could be improved by monitoring the activity of a
star over many years, but this is not usually practical. Instead, the
recent focus of activity-age relations has turned to chromospheric
emission in the form of the $R^{'}_{\rm HK}$ index.

\subsubsection{Chromospheric activity}

\begin{figure}
\begin{center}
\includegraphics[width=10cm]{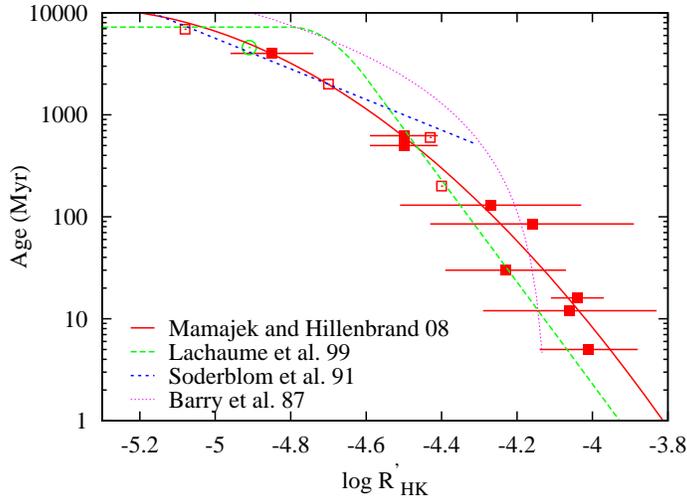}
\end{center}
\caption{The age-dependence of the chromospheric $R^{'}_{\rm HK}$
  index. The data are taken from Mamajek \& Hillenbrand (2008). Solid
  points are clusters with HK measurements on the Mt
  Wilson system. The horizontal bars represent the rms dispersion. Open symbols
  are several clusters with data recalibrated onto the Mt Wilson system
  (see Mamajek \& Hillenbrand for details); the Sun is also
  shown. Various age-activity relations proposed in the literature are
  shown. The Mamajek \& Hillenbrand curve was defined using the plotted
  data. Note the large dispersion in $R^{'}_{\rm HK}$ at ages $\leq
  200$\,Myr, presumably caused by a large spread in rotation rates.
} 
\label{rhkplot}
\end{figure}

In general as one considers activity indicators that arise from
lower/cooler layers in the magnetically heated outer atmosphere, the
time dependence becomes less steep ($\sim t^{-1/2}$, e.g. Skumanich
1972; Ribas \etal\ 2005; Findeisen,
Hillenbrand \& Soderblom 2011) and this appears to be true for G-, K-
and M-dwarfs (Stelzer \etal\ 2013). This is
because of the rather steep slope in the correlation between
chromospheric and coronal activity indicators (e.g. 
$L_X/L_{\rm bol} \propto (R^{\prime}_{\rm HK})^{3.46}$; Mamajek \&
Hillenbrand 2008).
However, typical levels of
variability in the $R^{\prime}_{\rm HK}$ index for stars that are young or old are only 5--10 per
cent (Baliunas \etal\ 1995), though young stars are still of course
affected by a spread in rotation rates and saturation of chromospheric
emission. So, even after taking account of the shallower 
$R^{\prime}_{\rm HK}$-age relation, we see
that chromospheric activity ages should suffer less than coronal
activity ages from uncertainties 
due to magnetic activity cycles and other variability in older stars.

There are several flavours of $R^{\prime}_{\rm HK}$-age relations in
the literature (e.g. Barry, Cromwell \& Hege 1987; 
Soderblom, Duncan \& Johnson 1991; Lachaume
\etal\ 1999), which are reviewed in some detail by Mamajek \&
Hillenbrand (2008), who then provide an updated calibration (see
Fig.~\ref{rhkplot})
and also provide
a $L_x/L_{\rm bol}$-age relationship that is bootstrapped from the
chromospheric calibration. The $R^{\prime}_{\rm HK}$-age relation suffers less
than the $L_x/L_{\rm bol}$-age relation from problems with variability
at older ages as discussed above, but the limitations associated with spreads in rotation rate and
hence activity at younger ages are just as severe. A further key advantage of
the $R^{\prime}_{\rm HK}$-age relation is that there are good data for
solar-type stars in
the old open clusters M67 and some data in NGC~188. These calibration ``points'', consisting of a
set of coeval stars at 4 and 6.9\,Gyr, 
give an excellent estimate of the precision of the
method. Further tests are provided by the comparison of chromospheric
ages for stars in wide binary systems, with either similar components
or components with different masses. 

\subsection{Problems, precision and accuracy}

 Mamajek \& Hillenbrand (2008) conclude that for solar-type
stars older than a few hundred Myr, carefully measured $R^{\prime}_{\rm
  HK}$ values yield $\log$ age to $\pm 0.2$~dex, or an age precision of
$\sim 60$ per cent. The uncertainty grows rapidly at younger ages, due
to the growing dispersion in $R^{\prime}_{\rm
  HK}$ (see Fig.~\ref{rhkplot}), to become unusable at $\leq 100$\,Myr 
except in providing an upper limit to the age of a star. For reasons that are not clear (perhaps binaries have smaller amplitude
activity cycles?),
the dispersion in empirically determined age between the components of
binary systems is lower than the dispersion implied by the spread of
observed chromospheric activity in the presumably coeval stars of the
Hyades and M67.

A further limitation of the chromospheric activity-age relation is that
calibrating data for lower mass stars is more sparse. Attempts to
determine the slope of the $R^{\prime}_{\rm  HK}$-mass (or colour)
relation from cluster data yield a wide diversity of results. Instead,
Mamajek \& Hillenbrand (2008) use a newly derived activity-$N_R$
relation and combine this with a gyrochronology relation to create an
activity-age-colour relationship, calibrated for F7-K2 main sequence
stars. This assumes that stars have
converged to the I-sequence and also makes the same assumptions about
the separable nature of the colour and time dependence of rotation rate
used in other gyrochronology relations. The new relation does reduce
the dispersion in ages estimated for the binary components with
different masses, but the dispersion estimated for stars in M67
remains stubbornly at the $\pm 0.2$~dex level.
The situation is similar for X-ray activity indicators, though likely to
be worse at older ages despite a steeper decline with age, 
because coronal X-ray variability is much greater than that of the chromosphere 
in older stars. 

Further limitations to the technique mirror those discussed for
gyrochronology in section~\ref{gyrolimitations}. Whilst differential
rotation should not be a problem, the limited convergence of rotation
rates onto the I-sequence may be partly responsible for the dispersion
in ages estimated for coeval stars. Like gyrochronology, activity-age
relationships should not be used for close binary systems where tidal
or other interactions may have affected the rotation rates of the
components. Finally, like gyrochronology, the activity-age
relationships (both coronal and chromospheric) are poorly calibrated at
ages older than, and masses lower than, the Sun. Attempts to improve
this situation are briefly reviewed in section~\ref{improvements}.

\section{Lithium depletion and age}
\label{sec4}

The ``ecology'' of lithium in the universe makes it a fascinating probe
of many physical processes, ranging from the big-bang to cosmic ray
spallation reactions and mixing in stellar interiors. $^7$Li is
produced during the first minutes of a standard big-bang (Wagoner,
Fowler \& Hoyle 1967) at a predicted
abundance (post-{\it Planck}) of $^7$Li/H$=4.89 ^{+0.41}_{-0.39}
\times 10^{-10}$, or A(Li)$=2.69^{+0.03}_{-0.04}$, on a logarithmic
scale where A(H)$=12$ (Coc, Uzan \& Vangioni 2013). This abundance is
significantly higher than seen in old population~II stars, which
exhibit a plateau of Li abundance versus $T_{\rm eff}$ at
A(Li)$=2.20\pm 0.09$ (Sbordone \etal\ 2010) -- the "Spite plateau''
(Spite \& Spite 1982). This discrepancy suggests either ``new'' physics
beyond the standard big-bang model or that physical processes have been
able to deplete Li from the photospheres of these old stars.

On the other hand, the abundance of Li found in meteorites is
A(Li)$=3.26 \pm 0.05$ (Asplund \etal\ 2009), which implies that the
interstellar medium in the Galaxy becomes Li-enriched with
time. Several production mechanisms are under investigation; inside AGB
stars, cosmic ray spallation, novae (Prantzos 2012). The photospheric
solar $^7$Li abundance is A(Li)$=1.05 \pm 0.10$ and observations of solar-type
stars in the field and open clusters reveal a wide dispersion of A(Li),
from less than the solar value to approximately the meteoritic
abundance, clearly indicating that depletion mechanisms are at work.
It is the time-dependence of these depletion processes that makes Li 
abundance a potential age indicator.

\subsection{Lithium in pre main sequence stars}

\subsubsection{The astrophysics of PMS Li depletion}
\label{astroli}

After PMS stars are born, they initially 
contract along fully-convective Hayashi tracks. Once
their cores reach temperatures of $\sim 3 \times10^{6}$~K, then Li
burning commences  through the reaction\footnote{The isotope $^6$Li is burned at lower temperatures
  and should be completely depleted in all the stars discussed
  here.}
$^7$Li(p,$^4$He)$^4$He.  The reaction is extremely temperature dependent $( \sim
T^{20}$; Bildsten \etal\ 1997),
the density dependence is secondary. Li-depleted material at the core is convectively
mixed upwards and replaced with fresh material and the star could then be
completely depleted of Li on a timescale of a few--10\,Myr (much less than the
Kelvin-Helmholtz contraction timescale).

In stars with $M<0.4\,M_{\odot}$, that remain fully convective right
through to the ZAMS, this is indeed what happens. However, higher mass stars
develop a central radiative core because their central opacity falls
far enough to reduce the temperature gradient below the critical value
necessary to trigger convection. Convective mixing to the core ceases
and the extent to which {\it photospheric} Li depletion will continue
depends now on the temperature of the convection zone base ($T_{\rm
  BCZ}$). In stars of $M \leq 0.6 M_{\odot}$ (based on the models of
Siess, Dufour \& Forestini 2000), $T_{\rm BCZ}$ remains above the Li-burning
threshold long enough to completely deplete Li in the photosphere, but
in more massive stars Li-depletion should eventually be halted as the
radiative core expands. If $M>1.1\,M_{\odot}$ the radiative core forms
before Li-depletion commences and such stars deplete very little Li.
$T_{\rm BCZ}$ is never hot enough to
allow Li-burning in stars with $M\geq 1.3\,M_{\odot}$ 
and their photospheric Li should remain at its initial
value.

It is worth emphasizing that the above discussion takes account only of
convective mixing of material and predicts that depletion of
photospheric Li should have ceased by $\sim 100$\,Myr in all stars with
$M\geq 0.4\,M_{\odot}$ and considerably earlier in stars of higher mass;
i.e. the pattern of Li depletion versus mass should be settled
prior to arrival on the ZAMS. Many flavours of evolutionary model have
made predictions about the onset and rate of photospheric Li depletion
and these can be used to define isochrones of Li depletion in the
A(Li) versus $T_{\rm eff}$ plane (see Fig.~\ref{pleiadesli}).

For stars that develop a radiative core, the predicted Li depletion
as a function of $T_{\rm eff}$ is highly sensitive to the physics
included in the models -- the opacities (and therefore metallicity), 
the efficiency of
convection parametrised in terms of a mixing length or overshooting, and the adopted atmospheres
(e.g. Chaboyer, Pinsonneault \& Demarque 1995b; Piau \& Turck-Chi\`eze 2002). For
instance, at $T_{\rm eff} \simeq 5000$\,K, the models of Baraffe
\etal\ (2002) with mixing lengths of 1.0 pressure scale height or 1.9
pressure scale heights (the value that matches the solar luminosity at
the solar age) have depleted Li by factors of 0.6 and 0.06
respectively at an age of 125\,Myr (see Fig.~\ref{pleiadesli}).

\subsubsection{Models vs observations}
\begin{figure}
\begin{center}
\includegraphics[width=10cm]{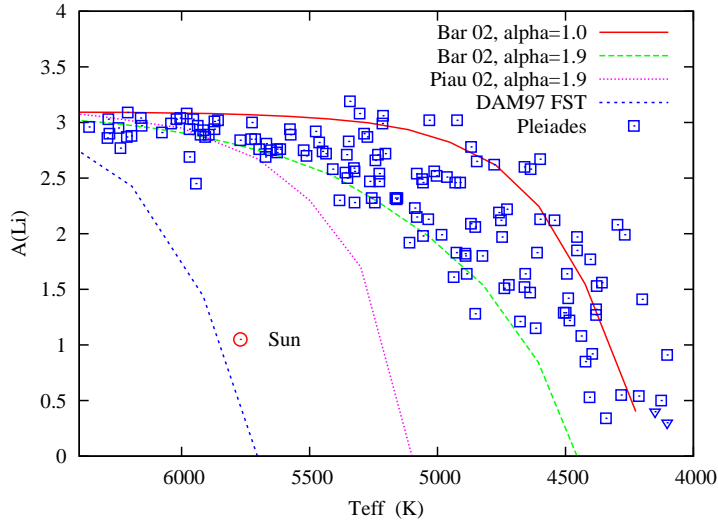}
\end{center}
\caption{Lithium abundances measured in the Pleiades (age 125\,Myr) at
  the end of PMS Li depletion (Soderblom et al. 1993; Jones
  \etal\ 1996). The lines are isochrones of Li depletion at 125\,Myr predicted by
  various models (D'Antona \& Mazzitelli 1997; Baraffe \etal\ 2002; Piau \& Turck-Chi\`eze 2002).
} 
\label{pleiadesli}
\end{figure}

Lithium is almost always measured using the resonance doublet of
Li\,{\sc i} at 6708\AA. There are other transitions in the optical spectrum
at 6104\AA\ and 8126\AA, but these are {\it much} weaker and blended
with stronger lines. 
Even in cool stars Li is almost completely
ionised, the line strengths are temperature sensitive (in warmer stars)
and subject to NLTE effects that perturb abundances by up to $0.3$\,dex
depending on the Li abundance, $T_{\rm eff}$ and metallicity of the
atmosphere (e.g. Carlsson \etal\ 1994). Basic curves of growth for the
Li\,{\sc i}\,6708\AA\ feature have been calculated by a number of authors
(e.g. Soderblom \etal\ 1993; Palla \etal\ 2007). These show that at
A(Li)$\sim 3$, the equivalent width (EW) is about 0.5\AA\ and in the
saturated part of the curve of growth in cool stars ($T_{\rm eff} <
3500\,K$), while it is weaker ($\sim 0.15$\AA), but more sensitive to
abundance, in solar
type stars. The 6708\AA\ line is also blended with a
Fe\,{\sc i} line at 6707.44\AA. This is much weaker than the Li feature
for stars with undepleted Li but a more accurate assessment of this
blend becomes important as Li is depleted (Soderblom \etal\ 1993).
Other problems associated with estimating an accurate Li abundance
arise from an accurate estimation of $T_{\rm eff}$, especially in
young, active stars with spots and chromospheres, and photospheric
veiling by an accretion continuum may need to be accounted for in PMS stars
with disks.

A further problem in comparing models with observations is that models
predict Li {\it depletion}, so an initial abundance must be
adopted. For most young clusters this is usually assumed to be close to
the solar meteoritic value; there is also evidence from very young
(assumed to be undepleted) T-Tauri stars that A(Li)$_{\rm init} \simeq
3.1$--3.4 (e.g. Mart\'in \etal\ 1994; Soderblom \etal\ 1999). There are
however plausible reasons from Galactic chemical evolution models and
some observational evidence that the initial Li may be positively
correlated with [Fe/H] (e.g. Ryan \etal\ 2001; Cummings \etal\ 2012). It
seems reasonable to assume that for young stars near the Sun there
could be a $\pm 0.1$--0.2 dex spread in the initial A(Li).

Figure~\ref{pleiadesli} represents the most basic comparison of PMS Li
depletion with models, showing Li abundances in the Pleiades, 
which has an age of 125\,Myr (Stauffer, Schulz \&
Kirkpatrick 1998), versus a number of
representative model isochrones (at $\simeq 100$\,Myr) 
from the literature (D'Antona \& Mazzitelli 1997;
Piau \& Turck-Chi\`eze 2002; Baraffe \etal\ 2002). The solar
photospheric Li abundance is also shown. This Figure illustrates
several important points:
\begin{itemize}
\item There appears to be little Li depletion (assuming an initial
  A(Li)$=3.2$) among G-stars and this is predicted by most of the
  models. As the scatter in abundance
  ($\simeq 0.2$ dex) is similar to the amount of depletion in G-stars
  and similar to the uncertainty in initial Li abundance, Li depletion
  will not be an accurate age indicator below 125\,Myr for these stars.
\item The models differ vastly in their predictions of PMS Li
  depletion in cooler stars. There are several differences between these models, but
  the dominant one as far as Li depletion is concerned is convective
  efficiency. 
\item Models that have a convective efficiency (mixing length) tuned to
  match the Sun's luminosity (the Piau \& Turck-Chi\`eze 2002 model and
  the Baraffe \etal\ 2002 models with mixing length set to 1.9 pressure
  scale heights) predict too much Li depletion. A lower convective
  efficiency provides a better match (see also Tognelli, Degl'Innocenti \&
  Prada Moroni 2012).
 \item A scatter in Li abundance develops in this coeval cluster at
   $T_{\rm eff}<5500$\,K that cannot be
   accounted for by observational uncertainties ($\sim 0.1$--0.2\,dex)
   or explained by the models shown.
\end{itemize}
The large disagreements between the various model flavours and the
failure of these models to match the Pleiades or to predict the scatter among
the K-stars means that Li abundance in solar-type stars (those that develop a
radiative core) cannot yet be used as anything but an
{\it empirical} age indicator. However, the scatter at a given $T_{\rm
  eff}$ (or colour) is a problem in that regard too, since stars of
similar age may show a wide dispersion in their Li abundances. Unless
the factors causing this dispersion can be identified, there is an
inevitable uncertainty on any age inferred from an Li abundance.

There is a long list of possible causes of the Li-abundance dispersion
that have been considered. It seems possible that some fraction of it
may be caused by atmospheric inhomogeneities or chromospheric heating
of the upper photosphere (e.g. Randich 2001; 
King \etal\ 2010), but it is unlikely to be
dominant given the lack of variability in the line strengths (Jeffries
1999) and the agreement between abundances derived from the 6104\AA\ and
6708\AA\ features. A big clue may be the correlation with rotation,
noted for the Pleiades by Soderblom et al. (1993) and 
Garc\'ia L\'opez, Rebolo \&
Mart\'in (1994),  and now seen in several other young
clusters (though not always so strongly e.g. IC~2602, Alpha Per and
several young kinematic groups --
Randich \etal\ 2001; da Silva \etal\ 2009; Balachandran, Mallik \& Lambert 2011), in the
sense that fast rotators (usually only projected rotation velocity is
available) have preserved more Li. Caution must be exercised in
interpreting such results unless spectral syntheses have been used to
estimate EWs or abundances, as line broadening and blending can cause
overestimated EWs from direct integration methods (Margheim 2007).
It is unlikely that the structural
effects of rotation have much influence, so attention has focused on
additional rotational mixing of Li, which may be greater in slower rotators
with more internal differential rotation (see Fig.~\ref{gallet13} and
Bouvier 2008; Eggenberger \etal\ 2012), 
or the inhibition of convective mixing by stronger
magnetic fields in rapid rotators (e.g. 
D'Antona, Ventura \& Mazzitelli 2000; Somers \& Pinsonneault 2014).

\begin{figure}
\begin{center}
\includegraphics[width=10cm]{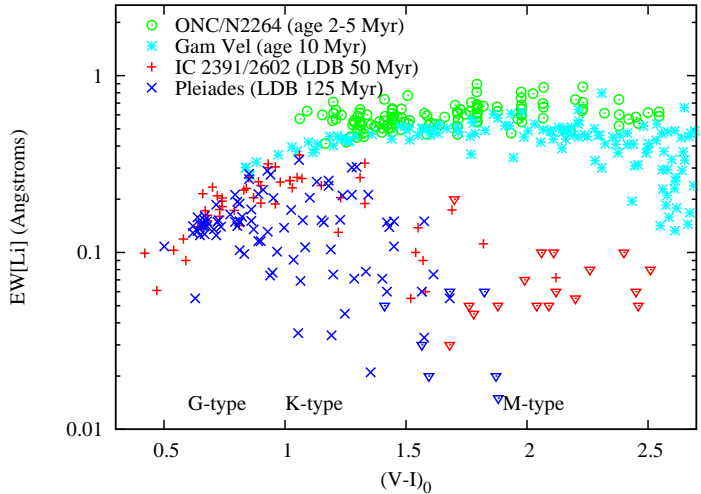}
\end{center}
\caption{Empirical Li depletion patterns for a set of fiducial clusters
  in the form of equivalent
  width of the 6708\AA\ features vs (dereddened) colour. Data are from
  Soderblom \etal\ (1993); Jones \etal\ (1996); Randich \etal\ (1997, 2001); Sergison \etal\ (2013); Jeffries \etal\ (2014). 
} 
\label{empiricalli}
\end{figure}

A further problem with using Li as an empirical age indicator in young
stars is that PMS Li depletion is predicted to be very sensitive to
metallicity. For example, the models of Piau \& Turck-Chi\`eze (2002)
show an order of magnitude increase in ZAMS Li depletion for a solar mass star
if the metallicity is increased by just 0.1 dex. The effect is smaller
at a fixed $T_{\rm eff}$ (about 0.2 dex in Li depletion per 0.1 dex
change in [M/H] at $T_{\rm eff} \simeq 5700$\,K -- Somers \&
Pinsonneault 2014), but grows towards cooler stars. Fortunately,
although puzzlingly for theoreticians, this extreme metallicity
dependence is not observed. Pairs of clusters with similar ages but
differing metallicities have only minor differences in Li depletion
pattern, 
and pairs of clusters with similar metallicities but different ages
have Li depletion patterns ordered
by age (e.g. Jeffries \etal\ 2002). It is possible that the metallicity
dependence is mostly masked by the time-dependence of 
processes that either inhibit or enhance PMS Li depletion (see Somers
\& Pinsonneault 2014). Hence an accurate knowledge of a {\it young} star's
metallicity does not greatly increase the precision with which an age can be
inferred from Li depletion.

Figure~\ref{empiricalli} shows how Li can be used to infer ages in the
form of a plot of Li~{\sc i}~6708\AA\ EWs versus colour (in the absence of
differential reddening, it is preferable to show the data in
the untransformed observational plane, rather than Li abundance versus
$T_{\rm eff}$) for a number of clusters with ages found by more certain
methods (e.g. see section~\ref{ldb}). Li is empirically sensitive to
age within a mass and age range where Li depletion has begun, but Li is
still detectable in the photosphere. If Li is undepleted then only an
upper limit to an age is possible, whilst if all the Li has gone then
an age lower limit is implied. For isolated stars this means that Li can
be used to estimate ages between about 10 and 50\,Myr for M-dwarfs and
between about 20 and a few 100\,Myr in K-dwarfs (Li has disappeared in
K-dwarfs by 600\,Myr in the Hyades -- Soderblom \etal\ 1995). The dispersion at a given
age, limits precision to about a factor of two.
G-dwarfs deplete little Li
on the PMS, and as this depletion is comparable to uncertainties
in initial Li, then Li depletion cannot be confidently used to estimate
ages in the range shown.

As a result of the above discussion, Li has mainly been used in the
literature for {\it identifying} young stars in circumstances where
their ages are otherwise uncertain (e.g. they cannot be placed
on an HR diagram because their distances are unknown). A boundary can
be defined in Fig.~\ref{empiricalli}, such that a star with an EW(Li)
above the boundary is younger than some desired threshold. Examples include
finding low-mass members of star forming regions, especially weak-lined
T-Tauri stars with no obvious accretion or circumstellar material 
(Alcala \etal\ 1996; Martin 1997 and many more since), or identifying
members of spatially dispersed members of young, kinematic groups
(e.g. Jeffries 1995; Zuckerman \& Webb 2000; Montes \etal\ 2001).
 Little effort has so far been applied to obtaining quantitative age
 estimates (or age probability distributions) for individual stars,
 though attempts have been made to put groups of coeval young stars in age
 order using Li (e.g. Mentuch \etal\ 2008). One notable problem in this
 endeavour is a lack of well-populated calibrating clusters between
 ages of 10 and 50\,Myr.

\subsection{Lithium in main sequence stars}

Figure~\ref{pleiadesli} shows that the Sun has depleted Li by $\sim
2$\,dex compared with similar mass stars at the ZAMS in the
Pleiades. Such depletion is entirely unanticipated by ``standard''
models that include only convective mixing and is also observed in field
stars at a range of $T_{\rm eff}$ around the solar value. Additional mixing
mechanisms have been proposed that will mix Li-depleted material from
the hot radiative core to the base of the convection zone and hence to the photosphere. These include
atomic diffusion (Richer \& Michaud 1993) or mixing induced by hydrodynamcal instabilities
associated with rotation or gravity waves
(e.g. Vauclair 1988; Garc\'ia L\'opez \& Spruit 1991; Pinsonneault 1997, 2010; Charbonnel \& Talon 2005). At present,
models of age-dependent Li depletion that incorporate 
these effects have significant uncertainties, including the usually
unknown rotational history of the star, and require tuning to
match the solar Li abundance and solar interior rotation profile
derived from
helioseismology (e.g. do Nascimento Jr \etal 2009). Furthermore, these extra
mixing mechanisms act in addition to standard PMS Li
depletion, but we have already seen that standard models
predict too much Li depletion in ZAMS clusters and fail to predict the
significant dispersion that is observed. Such uncertainties merely
prevent us from confidently inferring the age of a star by comparing its Li
abundance to a model. Indeed, the primary use of Li abundance data for
older stars and clusters has been to attempt to shed light on these
uncertain interior processes. However, 
the option is still open to empirically calibrate
Li depletion beyond the ZAMS using clusters, the Sun and other
stars of ``known'' age.

\subsubsection{Li in field stars}
General surveys of Li
abundances in field stars (e.g. Lambert \& Reddy 2004; 
Takeda \& Kawonomoto 2005; Ram\'irez \etal\ 2012) 
show a strong temperature dependence -- Li
is depleted by $\geq 3$\,dex or gone in all stars with $T_{\rm eff} <
5200\,K$, whilst stars at the solar $T_{\rm eff}$ show a $\sim 2$\,dex
dispersion, with the Sun towards the bottom of the distribution. At
higher temperatures the dispersion may narrow again, though there are
still some F-stars with very low abundances.  It is worth mentioning
that measuring the Li abundance in older stars is more
challenging, because the EW of the Li\,{\sc i}~6708\AA\ feature becomes
only a few m\AA\ at  A(Li)$\simeq 1$ in solar-type stars.

Naturally, one would like
to know to what extent this scatter is due to age (at a given $T_{\rm
  eff}$) and how much is due to confounding (but potentially resolvable)
parameters like metallicity (higher opacities lead to a deeper
convection zone, more PMS Li depletion and more effective MS mixing) 
or even the presence of planets (Bouvier 2008; Israelian
\etal\ 2009), and how much might be due to factors that
are more difficult to take into account. For example, the rotational history of the
star, which appears to affect PMS depletion and is predicted to be a
strong influence on MS Li depletion, is not easily determined once
rotational convergence has been reached ($\geq 500$\,Myr for solar-type
stars). 

There is considerable debate on these points in the literature. Baumann
\etal\ (2010) determine ages from HR diagrams (with typical uncertainties
of $\simeq \pm 1.5$\,Gyr) for solar analogues and find that there is
the expected correlation with age for stars in a tight $\pm 0.1$\,dex
[Fe/H] range around the solar value. There is still a A(Li) scatter of $\sim
1$\,dex at a given age, but they attribute this to spreads in the
initial rotation conditions of these stars (and find no evidence that
the presence of exoplanets is relevant). The larger sample of Ram\'irez
\etal\ (2012) also demonstrates that stars with $M<1.1\,M_{\odot}$ have greater Li
depletion with age and increasing metallicity (and again it is found
that exoplanet hosts do
not deplete more Li than similar stars with no detected planets), but
with a large scatter around the correlations.
On the other hand, for a small sample of stars with metallicity and mass very
close to the solar value, Monroe \etal\ (2013) claim an extremely tight
correlation between Li depletion and age, with essentially no scatter. 
The Sun's Li abundance lies on this
correlation and Monroe \etal\  suggest that previous studies, suggesting 
a large scatter in this relationship and that the solar A(Li)
was low, either had
insufficient spectral quality or encompassed too wide a range of
metallicity and mass to eliminate the dispersion caused by these
factors.
This latter study, which needs confirmation with a larger
sample, holds out the promise of a deterministic relation between A(Li)
and age if the mass and metallicity can be accurately
determined. However, it makes no reference to Li observations in older
clusters, which appear to tell a different story.

\subsubsection{Li in older post-ZAMS clusters}

The progress of post-ZAMS Li depletion  can be empirically followed in
the Li depletion patterns of older clusters. Samples in clusters should
be coeval (if membership can be established!) and have the added
advantage that a good mean metallicity can often be determined from the
analysis of a number of stellar spectra.  Initial studies included the
Hyades and Praesepe at ages of about 600\,Myr (e.g. Wallerstein, Herbig
\& Conti 1963; Boesgaard \& Tripicco 1986; Boesgaard \& Budge 1988;
Soderblom \etal\ 1990), NGC~752 (age 1.7\,Gyr, Hobbs \& Pilachowski
1986a) and M67 (age 4\,Gyr, Hobbs \& Pilachowski 1986b). These
studies, which were focused on stars at the solar $T_{\rm eff}$ and hotter,
immediately revealed what has been termed the ``F-star Li gap''. Stars
in a narrow range $6400< T_{\rm eff} < 6800$\,K can deplete their Li to
undetectable limits (A(Li)$<1.8$) by the age of the Hyades, a process
that appears to have begun at ages of $\sim 150$\,Myr (Steinhauer \&
Deliyannis 2004). The cause of the ``Li gap'' is still not fully
understood, but likely involves rotation-driven mixing (Deliyannis
\etal\ 1998). In principle, if $T_{\rm eff}$ can be precisely measured,
then the Li abundance in this temperature range could strongly
constrain the stellar age between 150\,Myr and $\sim 600$\,Myr.

Older stars of late F-type and cooler are fainter and harder to study
in distant clusters. Sestito \& Randich (2005) provide a review of the
important literature and a homogeneous reanalysis of the Li abundances. Randich (2010)
reviews subsequent observations, mostly made with the 8-m VLT. These
observations of solar-type stars in $\sim 10$ clusters with ages
between 600\,Myr and 8\,Gyr paint a confusing picture. There is little
scatter in the A(Li) vs $T_{\rm eff}$ relationship in the Hyades and
this seems to be true in some older
clusters like Be~32 and NGC~188 at ages of 6--8\,Gyr, but solar-type stars in these clusters have
10--20 times as much lithium as the Sun (Randich, Sestito \& Pallavicini
2003). On the other hand there are also examples of old clusters
(e.g. M67, NGC~6253) where there is a large dispersion in A(Li), with
some stars that are as depleted as the Sun, but others with
A(Li)$\simeq 2.3$ (e.g. Pasquini \etal\ 1997, 2008). 

It appears that the Li
in solar-type stars is slowly depleted (by a factor of 3--4) from about the age of
the Pleiades to 1\,Gyr. In this range it seems reasonable to use Li
as an age indicator -- the dispersion in A(Li) among clusters in this
interval suggests an age precision of only about 0.3~dex though. Beyond
1\,Gyr some stars appear to deplete Li further whilst others do not. It
is unclear at present what factors drive this dichotomy. If a star has
a very low abundance then it clearly indicates an age $>2$\,Gyr, but if
A(Li)$\sim 2$ then the constraint can only be that the age is $\geq 500$\,Myr.

\subsection{The lithium depletion boundary}

\label{ldb}

In stars that remain fully convective all the way to the ZAMS
($M<0.4\,M_{\odot}$) then Li burning will completely deplete Li from
the entire star. Core Li burning begins at an age which depends upon
the mass and hence luminosity of the PMS star. Li depletion
occurs quickly, so that in a group of coeval stars there should be a sharp
boundary between stars exhibiting complete Li depletion and stars with
only slightly lower luminosities that still retain all their initial Li
(Bildsten \etal\ 1997). This ``lithium depletion boundary'' (LDB) was first
used to confirm the identity of brown dwarf candidates in the Pleiades
-- substellar objects should have retained their Li in the Pleiades,
but older, more massive objects with similar spectral types would have
depleted Li (Basri, Marcy \& Graham 1996; Rebolo \etal\ 1996). Since then, the
LDB technique has been used to estimate the ages of 10 clusters and
associations by finding the luminosity (or absolute magnitude) of the
faintest star which has significantly depleted its Li.

The LDB method, as defined above, is almost independent of which
evolutionary models are used. The luminosity of the LDB is insensitive
to changes in the assumed convective efficiency, composition,
atmosphere and equation of state. Burke, Pinsonneault \& Sills (2004) performed a set
of tests using different input physics finding systematic uncertainties
in the range 3--8 per cent. It is worth noting though that ages might
be perturbed due to some factor that is assumed or ignored by all
models. An example could be extensive coverage by starspots; the
blocking of flux at the photosphere would inflate the star leading to a
lower central temperature and an underestimated LDB age (e.g. Jackson
\& Jeffries 2014; Somers \& Pinsonneault 2014).

The relationship between the luminosity of
the LDB and age is steep ($L_{\rm LDB} \propto t^{-2}$ at $20 <
t<100$\,Myr. As a result, typical
errors of 0.1~mag in distance moduli or bolometric corrections, lead to
10 per cent uncertainty in $L_{\rm LDB}$ and hence only 5 per cent age
uncertainties. Locating the LDB in relatively sparse datasets is
usually more of an issue, and the presence of unresolved binary systems,
which may appears 0.75 mag brighter than a single star of the same type,
can be a confusing factor (e.g. Jeffries \& Oliveira 2005).

The LDB method is only applicable to groups of stars in
clusters (see Table~1 in Soderblom \etal\ 2013), but has also been
applied to spatially dispersed members of young kinematic groups (the
Beta Pic group with an LDB age of $21\pm 4$\,Myr, Binks
\& Jeffries 2014; the Tuc-Hor group with an LDB age of $41\pm 2$\,Myr,
Kraus \etal\ 2014). In isolated very low-mass stars, the presence of undepleted Li at
a given luminosity can give an upper limit to the age, whilst if a
star has depleted all its Li then a lower limit to the age is implied.
Note that the above discussion refers only to the {\it luminosity} at
the LDB, which of course depends on a distance. The temperature or
spectral type at the LDB could be used as a distance-independent
marker, but unfortunately the model-insensitivity of the $L_{\rm
  LDB}$-age is not reproduced and the $T_{\rm LDB}$-age relation is
quite shallow. Hence such determinations are of much lower precision
and subject to significant model-dependent uncertainties dominated by
which atmospheres are used and the adopted convective efficiency in 
the models. A further limitation of the
LDB technique is that below 20\,Myr there is an increasing dispersion
in model predictions and below 10\,Myr some evolutionary models predict
no Li depletion at any mass. At older ages the $L_{\rm LDB}$-age
relation becomes much shallower and no Li depletion is expected in
objects with $M<0.065\,M_{\odot}$. However, the principal limitation is
telescope size. Objects around the LDB at ages $\geq 200$\,Myr
are so intrinsically faint that the $R\geq 3000$ spectroscopy
needed to measure the Li\,{\sc i}~6708\AA\ feature is impractical, even with 8-m telescopes.
 
Although the measurement of an LDB age is limited to only a few
clusters, the fact that the derived ages are mostly model-independent
means that they can be used to test or calibrate other age estimation
techniques at 20--200\,Myr in the same clusters (actually the oldest LDB age so far reported
is for Blanco 1 at $132\pm24$\,Myr; Cargile, James \& Jeffries
2010). So far, systematic comparisons have only been carried out between
LDB ages and ages determined by fitting the positions of high-mass
stars in the HR diagram (see Soderblom \etal\ 2013). Such comparisons reveal that high-mass models
without ``convective core overshoot'' yield ages that are 50 per cent
lower than the LDB ages in some clusters (e.g. the Pleiades and Alpha
Per clusters; Stauffer \etal\ 1998, 1999), implying that moderate core
overshooting or fast rotation in the high mass stars (or some
combination of the two) is required. Similar systematic comparison with
empirical age indicators, such as those discussed here, are likely to
be valuable additions that can improve the {\it accuracy} of the
empirical ages.

\section{The status of rotation, activity and lithium
  depletion as age indicators}

\label{sec5}

In this section I summarise the status of each of the considered age
indicators and point to ongoing developments that might improve the
precision and especially the accuracy of these ages. This review was
written from the point of view of estimating the ages of individual
stars in the field, which is likely to remain the most important
application -- e.g. estimating the ages of exoplanet hosts
(Walkowicz \& Basri 2013) or searching
for spatially dispersed members of kinematic groups (e.g. Shkolnik, Liu
\& Reid 2009; da Silva \etal\ 2009). New data from the {\it Gaia} satellite and large
spectroscopic surveys such as the {\it Gaia-ESO survey} (Gilmore
\etal\ 2012) will add impetus to this field, with the desire to
understand stellar ages and hence the chemical and dynamical history of our Galaxy.
All of the techniques discussed have limitations when applied to
isolated stars, caused by star-to-star dispersions in the
empirical relationships. Of course, if many (assumed) coeval stars are
available, then this dispersion can be overcome to give a mean age
estimate for the group. I have not emphasized this application because
usually in such cases there are age determination methods that are
higher up in the accuracy hierarchy (e.g. fitting cluster sequences in
an HR diagram, see section~1). However in the case of kinematic groups
where the distances to individual members may not be well known, then
the {\it distance-independence} of these empirical relationships may be
of benefit (e.g. Mentuch \etal\ 2008).

%\clearpage
\begin{figure}[ht!]
\begin{center}
\includegraphics[width=7.3cm]{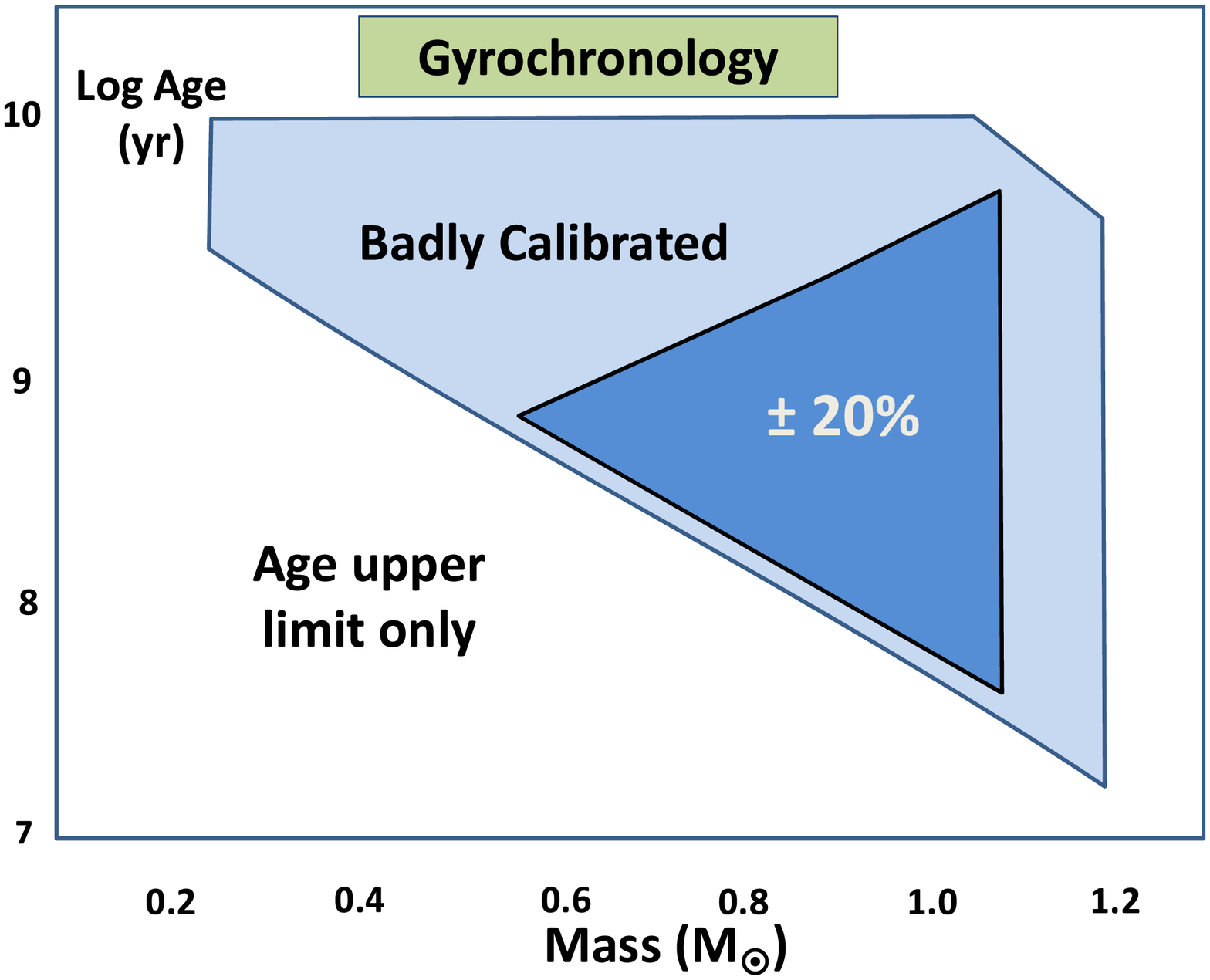}
\includegraphics[width=7.3cm]{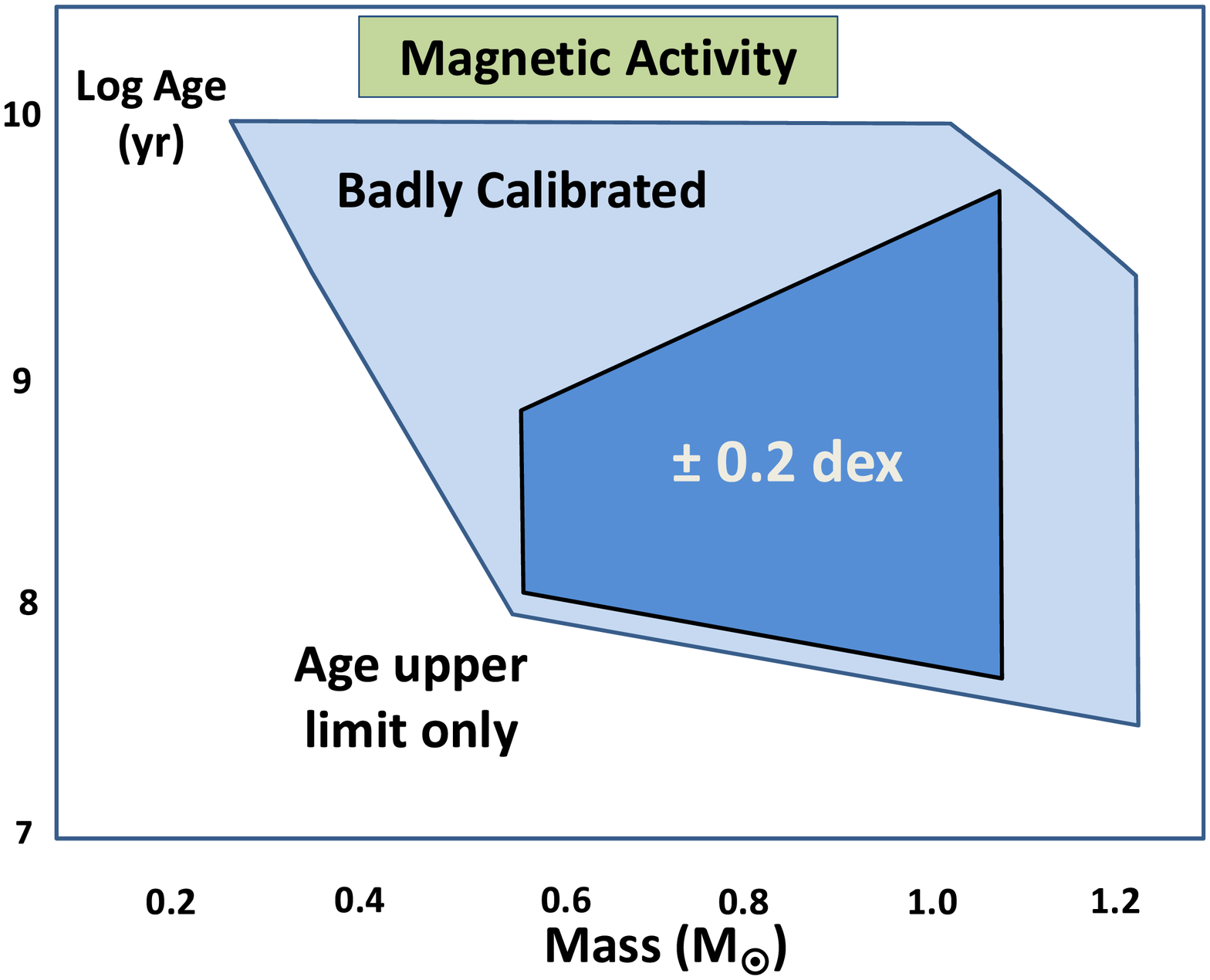}
\includegraphics[width=7.3cm]{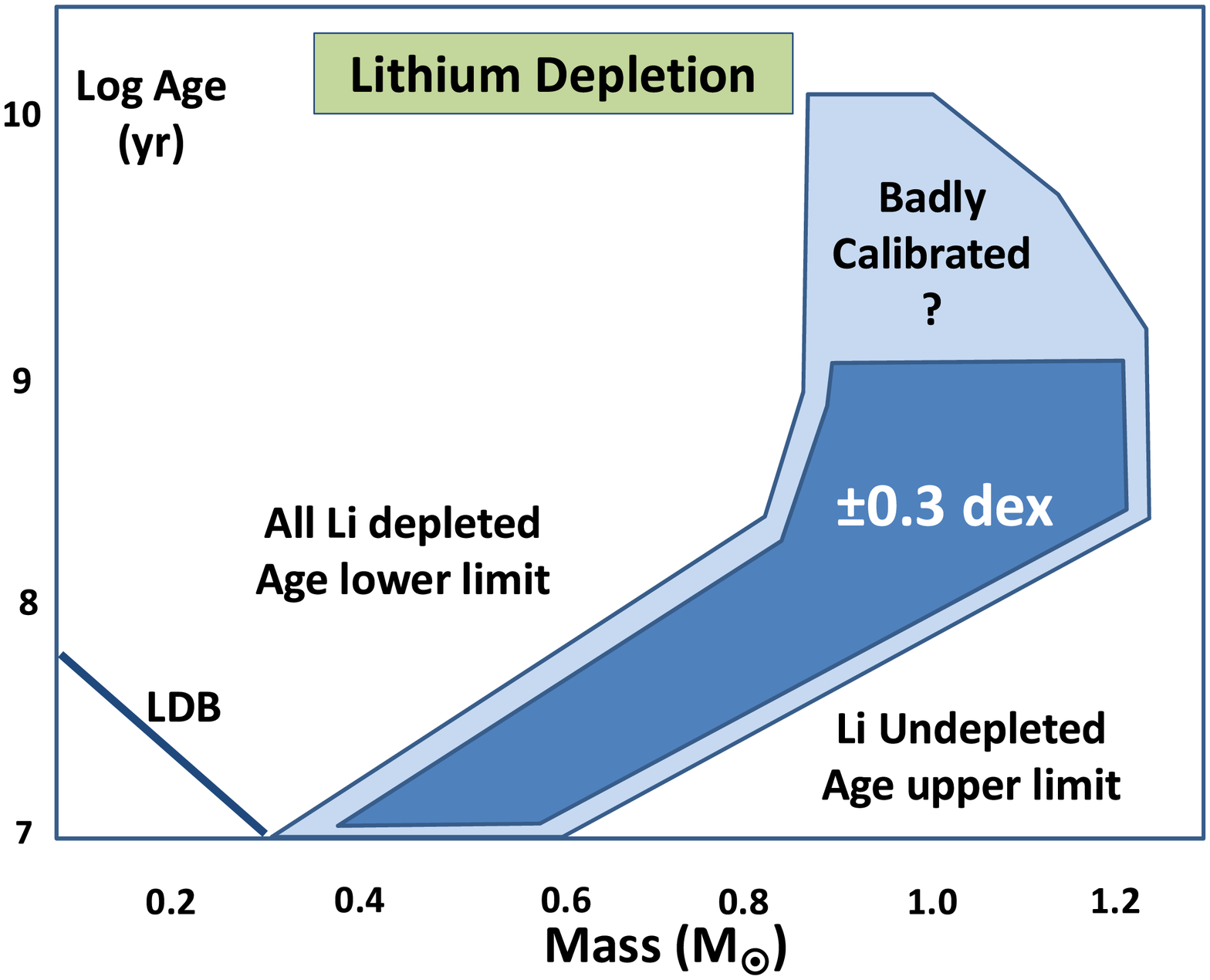}
\end{center}
\caption{Schematic diagrams illustrating the range of ages and masses
  where each empirical technique can be used to estimate ages or upper
  limits for individual stars.
} 
\label{summaryplot}
\end{figure}

\subsection{Applicability of the techniques}

Figure~\ref{summaryplot} summarises in a schematic way where each technique can feasibly
yield an age (or a limit to the age) as a function of age and mass.

\subsubsection{Rotation and gyrochronology}

Rotational evolution is not fully understood, however it appears that
magnetised winds act to cause the convergence of an initially wide
spread of rotation rates on a timescale of $\sim 100$\,Myr for G-type
stars, but as long as 1\,Gyr for stars of 0.5\,$M_{\odot}$, accounting
for the steeply sloped lower boundary in Fig.~\ref{summaryplot}a. Prior to
this convergence, stars have a dispersion in rotation rate and only age
upper limits can be determined. Once convergence is achieved (or nearly
achieved), then ages can be estimated with a precision of $\pm 20$ per
cent. The precision is determined in younger stars by the remaining
dispersion in rotation rates at a given age, and in older stars by
differential rotation with latitude. The precision will be much poorer
in stars with $M \leq 0.4\,M_{\odot}$, where convergence is likely to be
incomplete even at very old ages.

The dark shaded region of Fig.~\ref{summaryplot}a indicates that region
  where gyrochronology is well-calibrated using young clusters ($\leq
  1$\,Gyr) and the Sun. The lighter shaded region indicates where
  gyrochronology could work in principle, but where calibrating data
  are absent and so the accuracy of the technique may be poor when
  using calibrations extrapolated from younger and hotter stars.

\subsubsection{Magnetic activity}

As magnetic activity depends on rotation, then unsurprisingly, its
region of applicability, shown in Fig.~\ref{summaryplot}b, is similar
to that of gyrochronology. The lower boundary is determined by the large
spread of activity levels seen in young stars as a result of their
varied rotation rates. The lower boundary has a shallower slope than in
Fig.~\ref{summaryplot}a because at $0.5<M/M_{\odot}<1$ there should be
a small mass range at a given age, just below the mass at which
rotational convergence occurs, but where the {\it maximum} rotation
rate leads to activity lower than the saturated activity level and
hence an age could be estimated. However at very low masses, the rapid
increase in convective turnover time leads to stars at a wide range of
ages having rotation rates fast enough to cause saturated activity.

Activity diagnostics are generally easier to measure than rotation
periods, especially in older stars and so the well-calibrated region
for the activity-age relationship is larger than for gyrochronology,
with $R^{\prime}_{\rm HK}$ data for a couple of older clusters
providing more confidence in the calibration around a solar
mass. Nevertheless, these relationships are poorly constrained at lower
masses and the precision is roughly three times worse than
gyrochronology in mature stars. This is likely due to magnetic activity
cycles, so in principle could be mitigated by repeated observation on
long timescales.

\subsubsection{Lithium depletion}

Lithium abundances can only yield an age in the range where Li is {\it
  being} depleted. If the Li has gone, then a lower limit to the age
can be inferred; if Li is undepleted, only an upper limit can be
determined. The shape of the applicability region shown in
Fig.~\ref{summaryplot}c is a function of the mass-dependent timescale
for PMS Li depletion for stars with $M<1\,M_{\odot}$ and the observed
timescale of main-sequence depletion in older solar-type
stars. Precision is unlikely to be better than a factor of two until it
is fully understood why there is a dispersion of Li abundance in stars
of the same $T_{\rm eff}$ and age in calibrating clusters.

Beyond 1\,Gyr, both the theoretical and observational pictures are
confused. There appears to be a wide range of possible Li abundances
for stars like the Sun. This may be connected with their rotational
history, the presence of planets or some other factor; but for now it
seems that Li abundance cannot be used to estimate ages in older stars.

At the low mass side of Fig.~\ref{summaryplot}c, the narrow stripe
represents (schematically) the lithum depletion boundary (LDB). Fully
convective low-mass stars deplete their Li very rapidly (in a few
Myr). Thus in individual stars it would normally only be possible to
provide a one-sided limit on the stellar age. However, the power of the
LDB is that in a coeval group of stars with a range of masses, the transition
across this diagonal boundary will take place at an age-dependent mass
or luminosity, allowing the age of the ensemble to be determined
accurately.

\subsection{Ongoing efforts to improve calibrations}
\label{improvements}

For all three of the discussed empirical methods there is a need for more
calibration to improve the {\it accuracy} of the ages and assess their
precision. On the gyrochronological front, determining rotation periods
in stars of known age and at lower masses requires (a) relatively
nearby old clusters that still have a low-mass population, despite the
ongoing processes of energy equipartition, mass segregation and tidal
stripping, or other samples of stars with ``known'' ages; 
(b) space-based observations because spot
modulation amplitudes in older stars are very small. Meibom
\etal\ (2011a) discuss results for NGC~6811, one of 4 clusters between
0.5 and 9\,Gyr that are present in the Kepler field. NGC~6811 has an
age of 1~Gyr and a population spanning F- to early K-types. Of more
interest will be the results for NGC~6819 (2.5\,Gyr) and NGC~6791
(9\,Gyr), though their low-mass populations may be very sparse.

It is also worth noting that the clusters M35, Praesepe, the Hyades and the
Pleiades are possible targets for the Kepler K2 mission  during 2014/15 (Howell
\etal\ 2014). Whilst these clusters are not old, the data should
constrain much better the degree of rotational convergence between
125\,Myr and 600\,Myr, providing a much more complete census of
rotation periods, especially in the low-mass populations.

It is also possible to use Kepler data to provide asteroseismological
ages for many of the brighter (and predominantly solar-type) stars in
the Kepler field. Asteroseismology can give ages to perhaps 10--15 per
cent in these stars and they can then be used to calibrate rotation-age,
activity-age and Li-age relationships. This work has already begun:
Karoff \etal\ (2013) found $P = f(B-V)\, t^{0.81\pm0.10}$ for a small
  group of solar-type stars with asteroseismological ages and a
  Skumanich-like decay in chromospheric activity. 
From the Sun and a small sample of stars with $0.9<M/M_{\odot}<1.2$ and
ages from 1--9\,Gyr, Gar\'cia \etal\ (2014) determine a
  relationship $P \propto t^{0.52\pm0.09}$, in much closer accord with
  earlier work (see section~\ref{gyrouse}).

Other approaches to fix the ages of possible calibrating stars include
using objects which are in resolved binary systems with
either subgiants, giants or white dwarfs. The HR diagram (or white
dwarf cooling model)
is a much more precise tool for estimating the
companion age in these cases so, providing there is no possibility of previous
interaction or exchange of angular momentum, then the main sequence
companion could be used as a calibrator of empirical age
indicators. Examples include Silvestri \etal\ (2006), Chanam\'e \&
Ram\'irez (2012), Rebassa-Mansergas
\etal\ (2013) and Bochanski \etal\ (2013). No clear results have yet
emerged from these programs in terms of calibrating the activity-age or
rotation-age relationships.

\section{Summary}

The need for empirical methods of age estimation in low-mass stars is
likely to be present for some years to come. In this contribution I
have reviewed the astrophysical reasons that rotation, magnetic activity
and the photospheric abundance of lithium, change with time in low-mass
stars ($\leq 1.3\,M_{\odot}$). Whilst theoretical models that predict
these behaviours are improving rapidly, there are still very
significant uncertainties and semi-empirical components that prevent
their use in directly estimating stellar ages with any certainty, and
which require calibration using stars of known age.
Each of these empirical age indicators
can play a role in various domains of mass and age, that are
schematically illustrated in Fig.~\ref{summaryplot}. The rotation-age
relationship (or gyrochronology) offers the best prospect of
determining precise (to 20 per cent) 
ages in older stars, and could be complemented by
the use of PMS Li depletion to estimate the ages of younger stars at
low masses. Magnetic activity offers a less precise age determination
in older stars, but is usually easier to measure 
than rotation. In terms of accuracy, all these methods are
compromised to some extent by a lack of calibrating data in stars that
are older than the Sun or of lower masses than the Sun.

In very low mass stars, the sharp transition between stars that have
depleted all their lithium and stars with similar age but only slightly
lower luminosities that have preserved all their lithium (the lithium
depletion boundary), offers an almost model-independent way of
estimating an age for groups of coeval stars. This technique is
sensitive between ages of 20 and 200\,Myr and can be used to investigate
the uncertain physics in stellar models or calibrate empirical age
indicators.

\label{sec6}

\section*{Acknowledgements}

I would like to thank Corinne Charbonnel, Yveline Lebreton, David Valls-Gabaud and the rest of the local 
organising and scientific committees 
for arranging an exceptional meeting and for
inviting me to be a lecturer. Thanks are due to the Programme National
de Physique Stellaire and the CNRS for their financial support.

\nocite{bochanski13}
\nocite{silvestri06}
\nocite{rebassa13}
\nocite{chaname12}
\nocite{karoff13}
\nocite{garcia14}
\nocite{howell14}
\nocite{rhode01}
\nocite{dasilva09}
\nocite{shkolnik09}
\nocite{walkowicz13}
\nocite{stauffer99}
\nocite{cargile10}
\nocite{jackson14}
\nocite{kraus14}
\nocite{binks14}
\nocite{jeffries05}
\nocite{basri96}
\nocite{rebolo96}
\nocite{pasquini97}
\nocite{pasquini08}
\nocite{randich03}
\nocite{randich10}
\nocite{sestito05}
\nocite{deliyannis98}
\nocite{steinhauer04}
\nocite{vauclair88}
\nocite{garcialopez94}
\nocite{macgregor91}
\nocite{soderblom90}
\nocite{boesgaard86}
\nocite{wallerstein65}
\nocite{hobbs86a}
\nocite{boesgaard88}
\nocite{monroe13}
\nocite{israelian09}
\nocite{baumann10}
\nocite{vican12}
\nocite{charbonnel05}
\nocite{pinsonneault97}
\nocite{pinsonneault10}
\nocite{richer93}
\nocite{ramirez12}
\nocite{takeda05}
\nocite{lambert04}
\nocite{randich01b}
\nocite{randich97}
\nocite{mentuch08}
\nocite{soderblom95}
\nocite{montes01}
\nocite{zuckerman00}
\nocite{jeffries95}
\nocite{alcala96}
\nocite{martin97}
\nocite{jeffries02}
\nocite{stelzer13}
\nocite{findeisen11}
\nocite{huber11}
\nocite{stauffer98}
\nocite{somers14}
\nocite{dantona97}
\nocite{tognelli12}
\nocite{dantona00}
\nocite{eggenberger12}
\nocite{bouvier08}
\nocite{randich01}
\nocite{balachandran11}
\nocite{jeffries99}
\nocite{king10}
\nocite{jones96}
\nocite{ryan01}
\nocite{cummings12}
\nocite{soderblom99}
\nocite{martin94}
\nocite{palla07}
\nocite{soderblom93a}
\nocite{carlsson94}
\nocite{baraffe02}
\nocite{piau02}
\nocite{bildsten97}
\nocite{barry87}
\nocite{prantzos12}
\nocite{asplund09}
\nocite{spite82}
\nocite{sbordone10}
\nocite{wagoner67}
\nocite{coc13}
\nocite{lachaume99}
\nocite{soderblom91}
\nocite{mamajek08}
\nocite{baliunas95}
\nocite{robrade12}
\nocite{strong09}
\nocite{jeffries06}
\nocite{flaccomio12}
\nocite{wolk05}
\nocite{siess2000}
\nocite{flaccomio03}
\nocite{dobson89}
\nocite{vilhu87}
\nocite{wright11}
\nocite{jeffries11}
\nocite{pizzolato03}
\nocite{noyes84}
\nocite{pallavicini81}
\nocite{randich00}
\nocite{ribas05}
\nocite{telleschi05}
\nocite{gudel07}
\nocite{preibisch95}
\nocite{barnes10}
\nocite{brown14}
\nocite{barnes2003}
\nocite{irwin08}
\nocite{irwin11}
\nocite{mestelweiss87}
\nocite{reiners12}
\nocite{spada11}
\nocite{denissenkov10}
\nocite{matt12}
\nocite{eggenberger12}
\nocite{mathis13}
\nocite{moraux13}
\nocite{zanni13}
\nocite{matt05}
\nocite{koenigl91}
\nocite{cieza07}
\nocite{rebull06}
\nocite{prosser93}
\nocite{herbst02}
\nocite{irwinbouvier09}
\nocite{bouvier13b}
\nocite{gallet13}
\nocite{delorme11}
\nocite{barnes07}
\nocite{baliunas96}
\nocite{hartman10}
\nocite{hartman11}
\nocite{donahue96}
\nocite{meibom11a}
\nocite{mcquillan14}
\nocite{affer12}
\nocite{irwinm5009}
\nocite{herbst00}
\nocite{edwards93}
\nocite{camenzind90}
\nocite{hartman09}
\nocite{mcquillan13}
\nocite{kawaler88}
\nocite{mestelspruit87}
\nocite{chaboyer95}
\nocite{bouvier97}
\nocite{agueros11}
\nocite{meibom11b}
\nocite{radick87}
\nocite{krishnamurthi98}
\nocite{allain96}
\nocite{prosser93}
\nocite{reiners07}
\nocite{dravins90}
\nocite{gray76}
\nocite{kraft65}
\nocite{kraft67}
\nocite{skumanich72}
\nocite{soderblom10}
\nocite{soderblom13}
\nocite{perryman01}
\nocite{brown08}
\nocite{chaplin14}
\nocite{gai11}
\nocite{epstein14}
\nocite{chaboyer95b}
\nocite{meibom09}
\nocite{bouvier13rot}
\nocite{irwin07}
\nocite{makidon04}
\nocite{mathis13b}
\nocite{garcia91}
\nocite{charbonnel13}
\nocite{hobbs86b}
\nocite{sills00}
\nocite{simon98}
\nocite{favata08}
\nocite{collier09}
\nocite{reinhold13}
\nocite{burke04}
\nocite{gilmore12}
\nocite{sergison13}
\nocite{jeffries14}
\nocite{donascimento09}
\nocite{margheim07}

\bibliographystyle{astron}
\bibliography{roscoff13}

\begin{thebibliography}{}

\bibitem[\protect\astroncite{{Affer} et~al.}{2012}]{affer12}
{Affer}, L., {Micela}, G., {Favata}, F., and {Flaccomio}, E.: 2012,
\newblock {\em \mnras} {\bf 424}, 11

\bibitem[\protect\astroncite{{Ag{\"u}eros} et~al.}{2011}]{agueros11}
{Ag{\"u}eros}, M.~A., {Covey}, K.~R., {Lemonias}, J.~J., {Law}, N.~M., {Kraus},
  A., {Batalha}, N., {Bloom}, J.~S., {Cenko}, S.~B., {Kasliwal}, M.~M.,
  {Kulkarni}, S.~R., {Nugent}, P.~E., {Ofek}, E.~O., {Poznanski}, D., and
  {Quimby}, R.~M.: 2011,
\newblock {\em \apj} {\bf 740}, 110

\bibitem[\protect\astroncite{{Alcala} et~al.}{1996}]{alcala96}
{Alcala}, J.~M., {Terranegra}, L., {Wichmann}, R., {Chavarria-K.}, C.,
  {Krautter}, J., {Schmitt}, J.~H.~M.~M., {Moreno-Corral}, M.~A., {de Lara},
  E., and {Wagner}, R.~M.: 1996,
\newblock {\em \aaps} {\bf 119}, 7

\bibitem[\protect\astroncite{{Allain} et~al.}{1996}]{allain96}
{Allain}, S., {Bouvier}, J., {Prosser}, C., {Marschall}, L.~A., and
  {Laaksonen}, B.~D.: 1996,
\newblock {\em \aap} {\bf 305}, 498

\bibitem[\protect\astroncite{{Asplund} et~al.}{2009}]{asplund09}
{Asplund}, M., {Grevesse}, N., {Sauval}, A.~J., and {Scott}, P.: 2009,
\newblock {\em \araa} {\bf 47}, 481

\bibitem[\protect\astroncite{{Balachandran} et~al.}{2011}]{balachandran11}
{Balachandran}, S.~C., {Mallik}, S.~V., and {Lambert}, D.~L.: 2011,
\newblock {\em \mnras} {\bf 410}, 2526

\bibitem[\protect\astroncite{{Baliunas} et~al.}{1995}]{baliunas95}
{Baliunas}, S.~L., {Donahue}, R.~A., {Soon}, W.~H., {Horne}, J.~H., {Frazer},
  J., {Woodard-Eklund}, L., {Bradford}, M., {Rao}, L.~M., {Wilson}, O.~C.,
  {Zhang}, Q., {Bennett}, W., {Briggs}, J., {Carroll}, S.~M., {Duncan}, D.~K.,
  {Figueroa}, D., {Lanning}, H.~H., {Misch}, T., {Mueller}, J., {Noyes}, R.~W.,
  {Poppe}, D., {Porter}, A.~C., {Robinson}, C.~R., {Russell}, J., {Shelton},
  J.~C., {Soyumer}, T., {Vaughan}, A.~H., and {Whitney}, J.~H.: 1995,
\newblock {\em \apj} {\bf 438}, 269

\bibitem[\protect\astroncite{{Baliunas} et~al.}{1996}]{baliunas96}
{Baliunas}, S.~L., {Nesme-Ribes}, E., {Sokoloff}, D., and {Soon}, W.~H.: 1996,
\newblock {\em \apj} {\bf 460}, 848

\bibitem[\protect\astroncite{{Baraffe} et~al.}{2002}]{baraffe02}
{Baraffe}, I., {Chabrier}, G., {Allard}, F., and {Hauschildt}, P.~H.: 2002,
\newblock {\em \aap} {\bf 382}, 563

\bibitem[\protect\astroncite{{Barnes}}{2003}]{barnes2003}
{Barnes}, S.~A.: 2003,
\newblock {\em \apj} {\bf 586}, 464

\bibitem[\protect\astroncite{{Barnes}}{2007}]{barnes07}
{Barnes}, S.~A.: 2007,
\newblock {\em \apj} {\bf 669}, 1167

\bibitem[\protect\astroncite{{Barnes} and {Kim}}{2010}]{barnes10}
{Barnes}, S.~A. and {Kim}, Y.-C.: 2010,
\newblock {\em \apj} {\bf 721}, 675

\bibitem[\protect\astroncite{{Barry} et~al.}{1987}]{barry87}
{Barry}, D.~C., {Cromwell}, R.~H., and {Hege}, E.~K.: 1987,
\newblock {\em \apj} {\bf 315}, 264

\bibitem[\protect\astroncite{{Basri} et~al.}{1996}]{basri96}
{Basri}, G., {Marcy}, G.~W., and {Graham}, J.~R.: 1996,
\newblock {\em \apj} {\bf 458}, 600

\bibitem[\protect\astroncite{{Baumann} et~al.}{2010}]{baumann10}
{Baumann}, P., {Ram{\'{\i}}rez}, I., {Mel{\'e}ndez}, J., {Asplund}, M., and
  {Lind}, K.: 2010,
\newblock {\em \aap} {\bf 519}, A87

\bibitem[\protect\astroncite{{Bildsten} et~al.}{1997}]{bildsten97}
{Bildsten}, L., {Brown}, E.~F., {Matzner}, C.~D., and {Ushomirsky}, G.: 1997,
\newblock {\em \apj} {\bf 482}, 442

\bibitem[\protect\astroncite{{Binks} and {Jeffries}}{2014}]{binks14}
{Binks}, A.~S. and {Jeffries}, R.~D.: 2014,
\newblock {\em \mnras} {\bf 438}, L11

\bibitem[\protect\astroncite{{Bochanski} et~al.}{2013}]{bochanski13}
{Bochanski}, J.~J., {Hawley}, S.~L., {Covey}, K.~R., {Ag{\"u}eros}, M.~A.,
  {Baraffe}, I., {Catal{\'a}n}, S., {Mohanty}, S., {Rice}, E.~L., and {West},
  A.~A.: 2013,
\newblock {\em Astronomische Nachrichten} {\bf 334}, 44

\bibitem[\protect\astroncite{{Boesgaard} and {Budge}}{1988}]{boesgaard88}
{Boesgaard}, A.~M. and {Budge}, K.~G.: 1988,
\newblock {\em \apj} {\bf 332}, 410

\bibitem[\protect\astroncite{{Boesgaard} and {Tripicco}}{1986}]{boesgaard86}
{Boesgaard}, A.~M. and {Tripicco}, M.~J.: 1986,
\newblock {\em \apjl} {\bf 302}, L49

\bibitem[\protect\astroncite{{Bouvier}}{2008}]{bouvier08}
{Bouvier}, J.: 2008,
\newblock {\em \aap} {\bf 489}, L53

\bibitem[\protect\astroncite{{Bouvier}}{2013}]{bouvier13rot}
{Bouvier}, J.: 2013,
\newblock in {\em EAS Publications Series}, Vol.~62 of {\em EAS Publications
  Series}, pp 143--168

\bibitem[\protect\astroncite{{Bouvier} et~al.}{1997}]{bouvier97}
{Bouvier}, J., {Forestini}, M., and {Allain}, S.: 1997,
\newblock {\em \aap} {\bf 326}, 1023

\bibitem[\protect\astroncite{{Bouvier} et~al.}{2013}]{bouvier13b}
{Bouvier}, J., {Matt}, S.~P., {Mohanty}, S., {Scholz}, A., {Stassun}, K.~G.,
  and {Zanni}, C.: 2013,
\newblock {\em ArXiv e-prints 1309.7851}

\bibitem[\protect\astroncite{{Brown}}{2008}]{brown08}
{Brown}, A.~G.~A.: 2008,
\newblock in C.~A.~L. {Bailer-Jones} (ed.), {\em American Institute of Physics
  Conference Series}, Vol. 1082 of {\em American Institute of Physics
  Conference Series}, pp 209--215

\bibitem[\protect\astroncite{{Brown}}{2014}]{brown14}
{Brown}, T.~M.: 2014,
\newblock {\em ArXiv e-prints 1403.4525}

\bibitem[\protect\astroncite{{Burke} et~al.}{2004}]{burke04}
{Burke}, C.~J., {Pinsonneault}, M.~H., and {Sills}, A.: 2004,
\newblock {\em \apj} {\bf 604}, 272

\bibitem[\protect\astroncite{{Camenzind}}{1990}]{camenzind90}
{Camenzind}, M.: 1990,
\newblock in G. {Klare} (ed.), {\em Reviews in Modern Astronomy}, Vol.~3 of
  {\em Reviews in Modern Astronomy}, pp 234--265

\bibitem[\protect\astroncite{{Cargile} et~al.}{2010}]{cargile10}
{Cargile}, P.~A., {James}, D.~J., and {Jeffries}, R.~D.: 2010,
\newblock {\em \apjl} {\bf 725}, L111

\bibitem[\protect\astroncite{{Carlsson} et~al.}{1994}]{carlsson94}
{Carlsson}, M., {Rutten}, R.~J., {Bruls}, J.~H.~M.~J., and {Shchukina}, N.~G.:
  1994,
\newblock {\em \aap} {\bf 288}, 860

\bibitem[\protect\astroncite{{Chaboyer} et~al.}{1995a}]{chaboyer95}
{Chaboyer}, B., {Demarque}, P., and {Pinsonneault}, M.~H.: 1995a,
\newblock {\em \apj} {\bf 441}, 865

\bibitem[\protect\astroncite{{Chaboyer} et~al.}{1995b}]{chaboyer95b}
{Chaboyer}, B., {Demarque}, P., and {Pinsonneault}, M.~H.: 1995b,
\newblock {\em \apj} {\bf 441}, 876

\bibitem[\protect\astroncite{{Chanam{\'e}} and
  {Ram{\'{\i}}rez}}{2012}]{chaname12}
{Chanam{\'e}}, J. and {Ram{\'{\i}}rez}, I.: 2012,
\newblock {\em \apj} {\bf 746}, 102

\bibitem[\protect\astroncite{{Chaplin} et~al.}{2014}]{chaplin14}
{Chaplin}, W.~J., {Basu}, S., {Huber}, D., {Serenelli}, A., {Casagrande}, L.,
  {Silva Aguirre}, V., {Ball}, W.~H., {Creevey}, O.~L., {Gizon}, L.,
  {Handberg}, R., {Karoff}, C., {Lutz}, R., {Marques}, J.~P., {Miglio}, A.,
  {Stello}, D., {Suran}, M.~D., {Pricopi}, D., {Metcalfe}, T.~S., {Monteiro},
  M.~J.~P.~F.~G., {Molenda-{\.Z}akowicz}, J., {Appourchaux}, T.,
  {Christensen-Dalsgaard}, J., {Elsworth}, Y., {Garc{\'{\i}}a}, R.~A.,
  {Houdek}, G., {Kjeldsen}, H., {Bonanno}, A., {Campante}, T.~L., {Corsaro},
  E., {Gaulme}, P., {Hekker}, S., {Mathur}, S., {Mosser}, B., {R{\'e}gulo}, C.,
  and {Salabert}, D.: 2014,
\newblock {\em \apjs} {\bf 210}, 1

\bibitem[\protect\astroncite{{Charbonnel} et~al.}{2013}]{charbonnel13}
{Charbonnel}, C., {Decressin}, T., {Amard}, L., {Palacios}, A., and {Talon},
  S.: 2013,
\newblock {\em \aap} {\bf 554}, A40

\bibitem[\protect\astroncite{{Charbonnel} and {Talon}}{2005}]{charbonnel05}
{Charbonnel}, C. and {Talon}, S.: 2005,
\newblock {\em Science} {\bf 309}, 2189

\bibitem[\protect\astroncite{{Cieza} and {Baliber}}{2007}]{cieza07}
{Cieza}, L. and {Baliber}, N.: 2007,
\newblock {\em \apj} {\bf 671}, 605

\bibitem[\protect\astroncite{{Coc} et~al.}{2013}]{coc13}
{Coc}, A., {Uzan}, J.-P., and {Vangioni}, E.: 2013,
\newblock {\em ArXiv e-prints 1307.6955}

\bibitem[\protect\astroncite{{Collier Cameron} et~al.}{2009}]{collier09}
{Collier Cameron}, A., {Davidson}, V.~A., {Hebb}, L., {Skinner}, G.,
  {Anderson}, D.~R., {Christian}, D.~J., {Clarkson}, W.~I., {Enoch}, B.,
  {Irwin}, J., {Joshi}, Y., {Haswell}, C.~A., {Hellier}, C., {Horne}, K.~D.,
  {Kane}, S.~R., {Lister}, T.~A., {Maxted}, P.~F.~L., {Norton}, A.~J.,
  {Parley}, N., {Pollacco}, D., {Ryans}, R., {Scholz}, A., {Skillen}, I.,
  {Smalley}, B., {Street}, R.~A., {West}, R.~G., {Wilson}, D.~M., and
  {Wheatley}, P.~J.: 2009,
\newblock {\em \mnras} {\bf 400}, 451

\bibitem[\protect\astroncite{{Cummings} et~al.}{2012}]{cummings12}
{Cummings}, J.~D., {Deliyannis}, C.~P., {Anthony-Twarog}, B., {Twarog}, B., and
  {Maderak}, R.~M.: 2012,
\newblock {\em \aj} {\bf 144}, 137

\bibitem[\protect\astroncite{{da Silva} et~al.}{2009}]{dasilva09}
{da Silva}, L., {Torres}, C.~A.~O., {de La Reza}, R., {Quast}, G.~R., {Melo},
  C.~H.~F., and {Sterzik}, M.~F.: 2009,
\newblock {\em \aap} {\bf 508}, 833

\bibitem[\protect\astroncite{{D'Antona} and {Mazzitelli}}{1997}]{dantona97}
{D'Antona}, F. and {Mazzitelli}, I.: 1997,
\newblock {\em MmSAI} {\bf 68}, 807

\bibitem[\protect\astroncite{{D'Antona} et~al.}{2000}]{dantona00}
{D'Antona}, F., {Ventura}, P., and {Mazzitelli}, I.: 2000,
\newblock {\em \apjl} {\bf 543}, L77

\bibitem[\protect\astroncite{{Deliyannis} et~al.}{1998}]{deliyannis98}
{Deliyannis}, C.~P., {Boesgaard}, A.~M., {Stephens}, A., {King}, J.~R., {Vogt},
  S.~S., and {Keane}, M.~J.: 1998,
\newblock {\em \apjl} {\bf 498}, L147

\bibitem[\protect\astroncite{{Delorme} et~al.}{2011}]{delorme11}
{Delorme}, P., {Collier Cameron}, A., {Hebb}, L., {Rostron}, J., {Lister},
  T.~A., {Norton}, A.~J., {Pollacco}, D., and {West}, R.~G.: 2011,
\newblock {\em \mnras} {\bf 413}, 2218

\bibitem[\protect\astroncite{{Denissenkov} et~al.}{2010}]{denissenkov10}
{Denissenkov}, P.~A., {Pinsonneault}, M., {Terndrup}, D.~M., and {Newsham}, G.:
  2010,
\newblock {\em \apj} {\bf 716}, 1269

\bibitem[\protect\astroncite{{Do Nascimento} et~al.}{2009}]{donascimento09}
{Do Nascimento}, Jr., J.~D., {Castro}, M., {Mel{\'e}ndez}, J., {Bazot}, M.,
  {Th{\'e}ado}, S., {Porto de Mello}, G.~F., and {de Medeiros}, J.~R.: 2009,
\newblock {\em \aap} {\bf 501}, 687

\bibitem[\protect\astroncite{{Dobson} and {Radick}}{1989}]{dobson89}
{Dobson}, A.~K. and {Radick}, R.~R.: 1989,
\newblock {\em \apj} {\bf 344}, 907

\bibitem[\protect\astroncite{{Donahue} et~al.}{1996}]{donahue96}
{Donahue}, R.~A., {Saar}, S.~H., and {Baliunas}, S.~L.: 1996,
\newblock {\em \apj} {\bf 466}, 384

\bibitem[\protect\astroncite{{Dravins} et~al.}{1990}]{dravins90}
{Dravins}, D., {Lindegren}, L., and {Torkelsson}, U.: 1990,
\newblock {\em \aap} {\bf 237}, 137

\bibitem[\protect\astroncite{{Edwards} et~al.}{1993}]{edwards93}
{Edwards}, S., {Strom}, S.~E., {Hartigan}, P., {Strom}, K.~M., {Hillenbrand},
  L.~A., {Herbst}, W., {Attridge}, J., {Merrill}, K.~M., {Probst}, R., and
  {Gatley}, I.: 1993,
\newblock {\em \aj} {\bf 106}, 372

\bibitem[\protect\astroncite{{Eggenberger} et~al.}{2012}]{eggenberger12}
{Eggenberger}, P., {Haemmerl{\'e}}, L., {Meynet}, G., and {Maeder}, A.: 2012,
\newblock {\em \aap} {\bf 539}, A70

\bibitem[\protect\astroncite{{Epstein} and {Pinsonneault}}{2014}]{epstein14}
{Epstein}, C.~R. and {Pinsonneault}, M.~H.: 2014,
\newblock {\em \apj} {\bf 780}, 159

\bibitem[\protect\astroncite{{Favata} et~al.}{2008}]{favata08}
{Favata}, F., {Micela}, G., {Orlando}, S., {Schmitt}, J.~H.~M.~M., {Sciortino},
  S., and {Hall}, J.: 2008,
\newblock {\em \aap} {\bf 490}, 1121

\bibitem[\protect\astroncite{{Findeisen} et~al.}{2011}]{findeisen11}
{Findeisen}, K., {Hillenbrand}, L., and {Soderblom}, D.: 2011,
\newblock {\em \aj} {\bf 142}, 23

\bibitem[\protect\astroncite{{Flaccomio} et~al.}{2003}]{flaccomio03}
{Flaccomio}, E., {Micela}, G., and {Sciortino}, S.: 2003,
\newblock {\em \aap} {\bf 402}, 277

\bibitem[\protect\astroncite{{Flaccomio} et~al.}{2012}]{flaccomio12}
{Flaccomio}, E., {Micela}, G., and {Sciortino}, S.: 2012,
\newblock {\em \aap} {\bf 548}, A85

\bibitem[\protect\astroncite{{Gai} et~al.}{2011}]{gai11}
{Gai}, N., {Basu}, S., {Chaplin}, W.~J., and {Elsworth}, Y.: 2011,
\newblock {\em \apj} {\bf 730}, 63

\bibitem[\protect\astroncite{{Gallet} and {Bouvier}}{2013}]{gallet13}
{Gallet}, F. and {Bouvier}, J.: 2013,
\newblock {\em \aap} {\bf 556}, A36

\bibitem[\protect\astroncite{{Garcia} et~al.}{2014}]{garcia14}
{Garcia}, R.~A., {Ceillier}, T., {Salabert}, D., {Mathur}, S., {van Saders},
  J.~L., {Pinsonneault}, M., {Ballot}, J., {Beck}, P.~G., {Bloemen}, S.,
  {Campante}, T.~L., {Davies}, G.~R., {do Nascimento}, Jr., J.-D., {Mathis},
  S., {Metcalfe}, T.~S., {Nielsen}, M.~B., {Suarez}, J.~C., {Chaplin}, W.~J.,
  {Jimenez}, A., and {Karoff}, C.: 2014,
\newblock {\em ArXiv e-prints 1403.7155}

\bibitem[\protect\astroncite{{Garcia Lopez} et~al.}{1994}]{garcialopez94}
{Garcia Lopez}, R.~J., {Rebolo}, R., and {Martin}, E.~L.: 1994,
\newblock {\em \aap} {\bf 282}, 518

\bibitem[\protect\astroncite{{Garcia Lopez} and {Spruit}}{1991}]{garcia91}
{Garcia Lopez}, R.~J. and {Spruit}, H.~C.: 1991,
\newblock {\em \apj} {\bf 377}, 268

\bibitem[\protect\astroncite{{Gilmore} et~al.}{2012}]{gilmore12}
{Gilmore}, G., {Randich}, S., {Asplund}, M., {Binney}, J., {Bonifacio}, P.,
  {Drew}, J., {Feltzing}, S., {Ferguson}, A., {Jeffries}, R., {Micela}, G.,
  {Negueruela}, I., {Prusti}, T., {Rix}, H.-W., {Vallenari}, A., {Alfaro}, E.,
  {Allende-Prieto}, C., {Babusiaux}, C., {Bensby}, T., {Blomme}, R.,
  {Bragaglia}, A., {Flaccomio}, E., {Fran{\c c}ois}, P., {Irwin}, M.,
  {Koposov}, S., {Korn}, A., {Lanzafame}, A., {Pancino}, E., {Paunzen}, E.,
  {Recio-Blanco}, A., {Sacco}, G., {Smiljanic}, R., {Van Eck}, S., and
  {Walton}, N.: 2012,
\newblock {\em The Messenger} {\bf 147}, 25

\bibitem[\protect\astroncite{{Gray}}{1976}]{gray76}
{Gray}, D.~F.: 1976,
\newblock {\em {The observation and analysis of stellar photospheres}},
\newblock Wiley-Interscience, New York

\bibitem[\protect\astroncite{{G{\"u}del}}{2007}]{gudel07}
{G{\"u}del}, M.: 2007,
\newblock {\em Living Reviews in Solar Physics} {\bf 4}, 3

\bibitem[\protect\astroncite{{Hartman} et~al.}{2010}]{hartman10}
{Hartman}, J.~D., {Bakos}, G.~{\'A}., {Kov{\'a}cs}, G., and {Noyes}, R.~W.:
  2010,
\newblock {\em \mnras} {\bf 408}, 475

\bibitem[\protect\astroncite{{Hartman} et~al.}{2011}]{hartman11}
{Hartman}, J.~D., {Bakos}, G.~{\'A}., {Noyes}, R.~W., {Sip{\H o}cz}, B.,
  {Kov{\'a}cs}, G., {Mazeh}, T., {Shporer}, A., and {P{\'a}l}, A.: 2011,
\newblock {\em \aj} {\bf 141}, 166

\bibitem[\protect\astroncite{{Hartman} et~al.}{2009}]{hartman09}
{Hartman}, J.~D., {Gaudi}, B.~S., {Pinsonneault}, M.~H., {Stanek}, K.~Z.,
  {Holman}, M.~J., {McLeod}, B.~A., {Meibom}, S., {Barranco}, J.~A., and
  {Kalirai}, J.~S.: 2009,
\newblock {\em \apj} {\bf 691}, 342

\bibitem[\protect\astroncite{{Herbst} et~al.}{2002}]{herbst02}
{Herbst}, W., {Bailer-Jones}, C.~A.~L., {Mundt}, R., {Meisenheimer}, K., and
  {Wackermann}, R.: 2002,
\newblock {\em \aap} {\bf 396}, 513

\bibitem[\protect\astroncite{{Herbst} et~al.}{2000}]{herbst00}
{Herbst}, W., {Rhode}, K.~L., {Hillenbrand}, L.~A., and {Curran}, G.: 2000,
\newblock {\em \aj} {\bf 119}, 261

\bibitem[\protect\astroncite{{Hobbs} and {Pilachowski}}{1986a}]{hobbs86b}
{Hobbs}, L.~M. and {Pilachowski}, C.: 1986a,
\newblock {\em \apjl} {\bf 311}, L37

\bibitem[\protect\astroncite{{Hobbs} and {Pilachowski}}{1986b}]{hobbs86a}
{Hobbs}, L.~M. and {Pilachowski}, C.: 1986b,
\newblock {\em \apjl} {\bf 309}, L17

\bibitem[\protect\astroncite{{Howell} et~al.}{2014}]{howell14}
{Howell}, S.~B., {Sobeck}, C., {Haas}, M., {Still}, M., {Barclay}, T.,
  {Mullally}, F., {Troeltzsch}, J., {Aigrain}, S., {Bryson}, S.~T., {Caldwell},
  D., {Chaplin}, W.~J., {Cochran}, W.~D., {Huber}, D., {Marcy}, G.~W.,
  {Miglio}, A., {Najita}, J.~R., {Smith}, M., {Twicken}, J.~D., and {Fortney},
  J.~J.: 2014,
\newblock {\em ArXiv e-prints 1402.5163}

\bibitem[\protect\astroncite{{Huber} et~al.}{2011}]{huber11}
{Huber}, D., {Bedding}, T.~R., {Stello}, D., {Hekker}, S., {Mathur}, S.,
  {Mosser}, B., {Verner}, G.~A., {Bonanno}, A., {Buzasi}, D.~L., {Campante},
  T.~L., {Elsworth}, Y.~P., {Hale}, S.~J., {Kallinger}, T., {Silva Aguirre},
  V., {Chaplin}, W.~J., {De Ridder}, J., {Garc{\'{\i}}a}, R.~A., {Appourchaux},
  T., {Frandsen}, S., {Houdek}, G., {Molenda-{\.Z}akowicz}, J., {Monteiro},
  M.~J.~P.~F.~G., {Christensen-Dalsgaard}, J., {Gilliland}, R.~L., {Kawaler},
  S.~D., {Kjeldsen}, H., {Broomhall}, A.~M., {Corsaro}, E., {Salabert}, D.,
  {Sanderfer}, D.~T., {Seader}, S.~E., and {Smith}, J.~C.: 2011,
\newblock {\em \apj} {\bf 743}, 143

\bibitem[\protect\astroncite{{Irwin} et~al.}{2009}]{irwinm5009}
{Irwin}, J., {Aigrain}, S., {Bouvier}, J., {Hebb}, L., {Hodgkin}, S., {Irwin},
  M., and {Moraux}, E.: 2009,
\newblock {\em \mnras} {\bf 392}, 1456

\bibitem[\protect\astroncite{{Irwin} et~al.}{2011}]{irwin11}
{Irwin}, J., {Berta}, Z.~K., {Burke}, C.~J., {Charbonneau}, D., {Nutzman}, P.,
  {West}, A.~A., and {Falco}, E.~E.: 2011,
\newblock {\em \apj} {\bf 727}, 56

\bibitem[\protect\astroncite{{Irwin} and {Bouvier}}{2009}]{irwinbouvier09}
{Irwin}, J. and {Bouvier}, J.: 2009,
\newblock in E.~E. {Mamajek}, D.~R. {Soderblom}, and R.~F.~G. {Wyse} (eds.),
  {\em IAU Symposium}, Vol. 258, pp 363--374

\bibitem[\protect\astroncite{{Irwin} et~al.}{2008}]{irwin08}
{Irwin}, J., {Hodgkin}, S., {Aigrain}, S., {Bouvier}, J., {Hebb}, L., and
  {Moraux}, E.: 2008,
\newblock {\em \mnras} {\bf 383}, 1588

\bibitem[\protect\astroncite{{Irwin} et~al.}{2007}]{irwin07}
{Irwin}, J., {Hodgkin}, S., {Aigrain}, S., {Hebb}, L., {Bouvier}, J., {Clarke},
  C., {Moraux}, E., and {Bramich}, D.~M.: 2007,
\newblock {\em \mnras} {\bf 377}, 741

\bibitem[\protect\astroncite{{Israelian} et~al.}{2009}]{israelian09}
{Israelian}, G., {Delgado Mena}, E., {Santos}, N.~C., {Sousa}, S.~G., {Mayor},
  M., {Udry}, S., {Dom{\'{\i}}nguez Cerde{\~n}a}, C., {Rebolo}, R., and
  {Randich}, S.: 2009,
\newblock {\em \nat} {\bf 462}, 189

\bibitem[\protect\astroncite{{Jackson} and {Jeffries}}{2014}]{jackson14}
{Jackson}, R.~J. and {Jeffries}, R.~D.: 2014,
\newblock {\em ArXiv e-prints 1404.0683}

\bibitem[\protect\astroncite{{Jeffries}}{1995}]{jeffries95}
{Jeffries}, R.~D.: 1995,
\newblock {\em \mnras} {\bf 273}, 559

\bibitem[\protect\astroncite{{Jeffries}}{1999}]{jeffries99}
{Jeffries}, R.~D.: 1999,
\newblock {\em \mnras} {\bf 309}, 189

\bibitem[\protect\astroncite{{Jeffries} et~al.}{2006}]{jeffries06}
{Jeffries}, R.~D., {Evans}, P.~A., {Pye}, J.~P., and {Briggs}, K.~R.: 2006,
\newblock {\em \mnras} {\bf 367}, 781

\bibitem[\protect\astroncite{{Jeffries} et~al.}{2011}]{jeffries11}
{Jeffries}, R.~D., {Jackson}, R.~J., {Briggs}, K.~R., {Evans}, P.~A., and
  {Pye}, J.~P.: 2011,
\newblock {\em \mnras} {\bf 411}, 2099

\bibitem[\protect\astroncite{{Jeffries} et~al.}{2014}]{jeffries14}
{Jeffries}, R.~D., {Jackson}, R.~J., {Cottaar}, M., {Koposov}, S.~E.,
  {Lanzafame}, A.~C., {Meyer}, M.~R., {Prisinzano}, L., {Randich}, S., {Sacco},
  G.~G., {Brugaletta}, E., {Caramazza}, M., {Damiani}, F., {Franciosini}, E.,
  {Frasca}, A., {Gilmore}, G., {Feltzing}, S., {Micela}, G., {Alfaro}, E.,
  {Bensby}, T., {Pancino}, E., {Recio-Blanco}, A., {de Laverny}, P., {Lewis},
  J., {Magrini}, L., {Morbidelli}, L., {Costado}, M.~T., {Jofr{\'e}}, P.,
  {Klutsch}, A., {Lind}, K., and {Maiorca}, E.: 2014,
\newblock {\em \aap} {\bf 563}, A94

\bibitem[\protect\astroncite{{Jeffries} and {Oliveira}}{2005}]{jeffries05}
{Jeffries}, R.~D. and {Oliveira}, J.~M.: 2005,
\newblock {\em \mnras} {\bf 358}, 13

\bibitem[\protect\astroncite{{Jeffries} et~al.}{2002}]{jeffries02}
{Jeffries}, R.~D., {Totten}, E.~J., {Harmer}, S., and {Deliyannis}, C.~P.:
  2002,
\newblock {\em \mnras} {\bf 336}, 1109

\bibitem[\protect\astroncite{{Jones} et~al.}{1996}]{jones96}
{Jones}, B.~F., {Shetrone}, M., {Fischer}, D., and {Soderblom}, D.~R.: 1996,
\newblock {\em \aj} {\bf 112}, 186

\bibitem[\protect\astroncite{{Karoff} et~al.}{2013}]{karoff13}
{Karoff}, C., {Metcalfe}, T.~S., {Chaplin}, W.~J., {Frandsen}, S., {Grundahl},
  F., {Kjeldsen}, H., {Christensen-Dalsgaard}, J., {Nielsen}, M.~B., {Frimann},
  S., {Thygesen}, A.~O., {Arentoft}, T., {Amby}, T.~M., {Sousa}, S.~G., and
  {Buzasi}, D.~L.: 2013,
\newblock {\em \mnras} {\bf 433}, 3227

\bibitem[\protect\astroncite{{Kawaler}}{1988}]{kawaler88}
{Kawaler}, S.~D.: 1988,
\newblock {\em \apj} {\bf 333}, 236

\bibitem[\protect\astroncite{{King} et~al.}{2010}]{king10}
{King}, J.~R., {Schuler}, S.~C., {Hobbs}, L.~M., and {Pinsonneault}, M.~H.:
  2010,
\newblock {\em \apj} {\bf 710}, 1610

\bibitem[\protect\astroncite{{Koenigl}}{1991}]{koenigl91}
{Koenigl}, A.: 1991,
\newblock {\em \apjl} {\bf 370}, L39

\bibitem[\protect\astroncite{{Kraft}}{1967}]{kraft67}
{Kraft}, R.~P.: 1967,
\newblock {\em \apj} {\bf 150}, 551

\bibitem[\protect\astroncite{{Kraft} and {Wilson}}{1965}]{kraft65}
{Kraft}, R.~P. and {Wilson}, O.~C.: 1965,
\newblock {\em \apj} {\bf 141}, 828

\bibitem[\protect\astroncite{{Kraus} et~al.}{2014}]{kraus14}
{Kraus}, A.~L., {Shkolnik}, E.~L., {Allers}, K.~N., and {Liu}, M.~C.: 2014,
\newblock {\em \aj} {\bf 147}, 146

\bibitem[\protect\astroncite{{Krishnamurthi} et~al.}{1998}]{krishnamurthi98}
{Krishnamurthi}, A., {Terndrup}, D.~M., {Pinsonneault}, M.~H., {Sellgren}, K.,
  {Stauffer}, J.~R., {Schild}, R., {Backman}, D.~E., {Beisser}, K.~B.,
  {Dahari}, D.~B., {Dasgupta}, A., {Hagelgans}, J.~T., {Seeds}, M.~A., {Anand},
  R., {Laaksonen}, B.~D., {Marschall}, L.~A., and {Ramseyer}, T.: 1998,
\newblock {\em \apj} {\bf 493}, 914

\bibitem[\protect\astroncite{{Lachaume} et~al.}{1999}]{lachaume99}
{Lachaume}, R., {Dominik}, C., {Lanz}, T., and {Habing}, H.~J.: 1999,
\newblock {\em \aap} {\bf 348}, 897

\bibitem[\protect\astroncite{{Lambert} and {Reddy}}{2004}]{lambert04}
{Lambert}, D.~L. and {Reddy}, B.~E.: 2004,
\newblock {\em \mnras} {\bf 349}, 757

\bibitem[\protect\astroncite{{MacGregor} and {Brenner}}{1991}]{macgregor91}
{MacGregor}, K.~B. and {Brenner}, M.: 1991,
\newblock {\em \apj} {\bf 376}, 204

\bibitem[\protect\astroncite{{Makidon} et~al.}{2004}]{makidon04}
{Makidon}, R.~B., {Rebull}, L.~M., {Strom}, S.~E., {Adams}, M.~T., and
  {Patten}, B.~M.: 2004,
\newblock {\em \aj} {\bf 127}, 2228

\bibitem[\protect\astroncite{{Mamajek} and {Hillenbrand}}{2008}]{mamajek08}
{Mamajek}, E.~E. and {Hillenbrand}, L.~A.: 2008,
\newblock {\em \apj} {\bf 687}, 1264

\bibitem[\protect\astroncite{{Margheim}}{2007}]{margheim07}
{Margheim}, S.~J.: 2007,
\newblock {\em Ph.D. thesis}, Indiana University

\bibitem[\protect\astroncite{{Martin}}{1997}]{martin97}
{Martin}, E.~L.: 1997,
\newblock {\em \aap} {\bf 321}, 492

\bibitem[\protect\astroncite{{Martin} et~al.}{1994}]{martin94}
{Martin}, E.~L., {Rebolo}, R., {Magazzu}, A., and {Pavlenko}, Y.~V.: 1994,
\newblock {\em \aap} {\bf 282}, 503

\bibitem[\protect\astroncite{{Mathis}}{2013}]{mathis13b}
{Mathis}, S.: 2013,
\newblock in {\em EAS Publications Series}, Vol.~63 of {\em EAS Publications
  Series}, pp 269--284

\bibitem[\protect\astroncite{{Mathis} et~al.}{2013}]{mathis13}
{Mathis}, S., {Decressin}, T., {Eggenberger}, P., and {Charbonnel}, C.: 2013,
\newblock {\em \aap} {\bf 558}, A11

\bibitem[\protect\astroncite{{Matt} and {Pudritz}}{2005}]{matt05}
{Matt}, S. and {Pudritz}, R.~E.: 2005,
\newblock {\em \apjl} {\bf 632}, L135

\bibitem[\protect\astroncite{{Matt} et~al.}{2012}]{matt12}
{Matt}, S.~P., {MacGregor}, K.~B., {Pinsonneault}, M.~H., and {Greene}, T.~P.:
  2012,
\newblock {\em \apjl} {\bf 754}, L26

\bibitem[\protect\astroncite{{McQuillan} et~al.}{2013}]{mcquillan13}
{McQuillan}, A., {Aigrain}, S., and {Mazeh}, T.: 2013,
\newblock {\em \mnras} {\bf 432}, 1203

\bibitem[\protect\astroncite{{McQuillan} et~al.}{2014}]{mcquillan14}
{McQuillan}, A., {Mazeh}, T., and {Aigrain}, S.: 2014,
\newblock {\em \apjs} {\bf 211}, 24

\bibitem[\protect\astroncite{{Meibom} et~al.}{2011a}]{meibom11a}
{Meibom}, S., {Barnes}, S.~A., {Latham}, D.~W., {Batalha}, N., {Borucki},
  W.~J., {Koch}, D.~G., {Basri}, G., {Walkowicz}, L.~M., {Janes}, K.~A.,
  {Jenkins}, J., {Van Cleve}, J., {Haas}, M.~R., {Bryson}, S.~T., {Dupree},
  A.~K., {Furesz}, G., {Szentgyorgyi}, A.~H., {Buchhave}, L.~A., {Clarke},
  B.~D., {Twicken}, J.~D., and {Quintana}, E.~V.: 2011a,
\newblock {\em \apjl} {\bf 733}, L9

\bibitem[\protect\astroncite{{Meibom} et~al.}{2009}]{meibom09}
{Meibom}, S., {Mathieu}, R.~D., and {Stassun}, K.~G.: 2009,
\newblock {\em \apj} {\bf 695}, 679

\bibitem[\protect\astroncite{{Meibom} et~al.}{2011b}]{meibom11b}
{Meibom}, S., {Mathieu}, R.~D., {Stassun}, K.~G., {Liebesny}, P., and {Saar},
  S.~H.: 2011b,
\newblock {\em \apj} {\bf 733}, 115

\bibitem[\protect\astroncite{{Mentuch} et~al.}{2008}]{mentuch08}
{Mentuch}, E., {Brandeker}, A., {van Kerkwijk}, M.~H., {Jayawardhana}, R., and
  {Hauschildt}, P.~H.: 2008,
\newblock {\em \apj} {\bf 689}, 1127

\bibitem[\protect\astroncite{{Mestel} and {Spruit}}{1987}]{mestelspruit87}
{Mestel}, L. and {Spruit}, H.~C.: 1987,
\newblock {\em \mnras} {\bf 226}, 57

\bibitem[\protect\astroncite{{Mestel} and {Weiss}}{1987}]{mestelweiss87}
{Mestel}, L. and {Weiss}, N.~O.: 1987,
\newblock {\em \mnras} {\bf 226}, 123

\bibitem[\protect\astroncite{{Monroe} et~al.}{2013}]{monroe13}
{Monroe}, T.~R., {Mel{\'e}ndez}, J., {Ram{\'{\i}}rez}, I., {Yong}, D.,
  {Bergemann}, M., {Asplund}, M., {Bedell}, M., {Tucci Maia}, M., {Bean}, J.,
  {Lind}, K., {Alves-Brito}, A., {Casagrande}, L., {Castro}, M., {do
  Nascimento}, J.-D., {Bazot}, M., and {Freitas}, F.~C.: 2013,
\newblock {\em \apjl} {\bf 774}, L32

\bibitem[\protect\astroncite{{Montes} et~al.}{2001}]{montes01}
{Montes}, D., {L{\'o}pez-Santiago}, J., {Fern{\'a}ndez-Figueroa}, M.~J., and
  {G{\'a}lvez}, M.~C.: 2001,
\newblock {\em \aap} {\bf 379}, 976

\bibitem[\protect\astroncite{{Moraux} et~al.}{2013}]{moraux13}
{Moraux}, E., {Artemenko}, S., {Bouvier}, J., {Irwin}, J., {Ibrahimov}, M.,
  {Magakian}, T., {Grankin}, K., {Nikogossian}, E., {Cardoso}, C., {Hodgkin},
  S., {Aigrain}, S., and {Movsessian}, T.~A.: 2013,
\newblock {\em \aap} {\bf 560}, A13

\bibitem[\protect\astroncite{{Noyes} et~al.}{1984}]{noyes84}
{Noyes}, R.~W., {Hartmann}, L.~W., {Baliunas}, S.~L., {Duncan}, D.~K., and
  {Vaughan}, A.~H.: 1984,
\newblock {\em \apj} {\bf 279}, 763

\bibitem[\protect\astroncite{{Palla} et~al.}{2007}]{palla07}
{Palla}, F., {Randich}, S., {Pavlenko}, Y.~V., {Flaccomio}, E., and
  {Pallavicini}, R.: 2007,
\newblock {\em \apjl} {\bf 659}, L41

\bibitem[\protect\astroncite{{Pallavicini} et~al.}{1981}]{pallavicini81}
{Pallavicini}, R., {Golub}, L., {Rosner}, R., {Vaiana}, G.~S., {Ayres}, T., and
  {Linsky}, J.~L.: 1981,
\newblock {\em \apj} {\bf 248}, 279

\bibitem[\protect\astroncite{{Pasquini} et~al.}{2008}]{pasquini08}
{Pasquini}, L., {Biazzo}, K., {Bonifacio}, P., {Randich}, S., and {Bedin},
  L.~R.: 2008,
\newblock {\em \aap} {\bf 489}, 677

\bibitem[\protect\astroncite{{Pasquini} et~al.}{1997}]{pasquini97}
{Pasquini}, L., {Randich}, S., and {Pallavicini}, R.: 1997,
\newblock {\em \aap} {\bf 325}, 535

\bibitem[\protect\astroncite{{Perryman} et~al.}{2001}]{perryman01}
{Perryman}, M.~A.~C., {de Boer}, K.~S., {Gilmore}, G., {H{\o}g}, E.,
  {Lattanzi}, M.~G., {Lindegren}, L., {Luri}, X., {Mignard}, F., {Pace}, O.,
  and {de Zeeuw}, P.~T.: 2001,
\newblock {\em \aap} {\bf 369}, 339

\bibitem[\protect\astroncite{{Piau} and {Turck-Chi{\`e}ze}}{2002}]{piau02}
{Piau}, L. and {Turck-Chi{\`e}ze}, S.: 2002,
\newblock {\em \apj} {\bf 566}, 419

\bibitem[\protect\astroncite{{Pinsonneault}}{1997}]{pinsonneault97}
{Pinsonneault}, M.: 1997,
\newblock {\em \araa} {\bf 35}, 557

\bibitem[\protect\astroncite{{Pinsonneault}}{2010}]{pinsonneault10}
{Pinsonneault}, M.~H.: 2010,
\newblock in C. {Charbonnel}, M. {Tosi}, F. {Primas}, and C. {Chiappini}
  (eds.), {\em IAU Symposium}, Vol. 268, pp 375--380

\bibitem[\protect\astroncite{{Pizzolato} et~al.}{2003}]{pizzolato03}
{Pizzolato}, N., {Maggio}, A., {Micela}, G., {Sciortino}, S., and {Ventura},
  P.: 2003,
\newblock {\em \aap} {\bf 397}, 147

\bibitem[\protect\astroncite{{Prantzos}}{2012}]{prantzos12}
{Prantzos}, N.: 2012,
\newblock {\em \aap} {\bf 542}, A67

\bibitem[\protect\astroncite{{Preibisch} and {Feigelson}}{2005}]{preibisch95}
{Preibisch}, T. and {Feigelson}, E.~D.: 2005,
\newblock {\em \apjs} {\bf 160}, 390

\bibitem[\protect\astroncite{{Prosser} et~al.}{1993}]{prosser93}
{Prosser}, C.~F., {Shetrone}, M.~D., {Marilli}, E., {Catalano}, S., {Williams},
  S.~D., {Backman}, D.~E., {Laaksonen}, B.~D., {Adige}, V., {Marschall}, L.~A.,
  and {Stauffer}, J.~R.: 1993,
\newblock {\em \pasp} {\bf 105}, 1407

\bibitem[\protect\astroncite{{Radick} et~al.}{1987}]{radick87}
{Radick}, R.~R., {Thompson}, D.~T., {Lockwood}, G.~W., {Duncan}, D.~K., and
  {Baggett}, W.~E.: 1987,
\newblock {\em \apj} {\bf 321}, 459

\bibitem[\protect\astroncite{{Ram{\'{\i}}rez} et~al.}{2012}]{ramirez12}
{Ram{\'{\i}}rez}, I., {Fish}, J.~R., {Lambert}, D.~L., and {Allende Prieto},
  C.: 2012,
\newblock {\em \apj} {\bf 756}, 46

\bibitem[\protect\astroncite{{Randich}}{2000}]{randich00}
{Randich}, S.: 2000,
\newblock in R. {Pallavicini}, G. {Micela}, and S. {Sciortino} (eds.), {\em
  Stellar Clusters and Associations: Convection, Rotation, and Dynamos}, Vol.
  198 of {\em Astronomical Society of the Pacific Conference Series}, p. 401

\bibitem[\protect\astroncite{{Randich}}{2001}]{randich01b}
{Randich}, S.: 2001,
\newblock {\em \aap} {\bf 377}, 512

\bibitem[\protect\astroncite{{Randich}}{2010}]{randich10}
{Randich}, S.: 2010,
\newblock in C. {Charbonnel}, M. {Tosi}, F. {Primas}, and C. {Chiappini}
  (eds.), {\em IAU Symposium}, Vol. 268 of {\em IAU Symposium}, pp 275--283

\bibitem[\protect\astroncite{{Randich} et~al.}{1997}]{randich97}
{Randich}, S., {Aharpour}, N., {Pallavicini}, R., {Prosser}, C.~F., and
  {Stauffer}, J.~R.: 1997,
\newblock {\em \aap} {\bf 323}, 86

\bibitem[\protect\astroncite{{Randich} et~al.}{2001}]{randich01}
{Randich}, S., {Pallavicini}, R., {Meola}, G., {Stauffer}, J.~R., and
  {Balachandran}, S.~C.: 2001,
\newblock {\em \aap} {\bf 372}, 862

\bibitem[\protect\astroncite{{Randich} et~al.}{2003}]{randich03}
{Randich}, S., {Sestito}, P., and {Pallavicini}, R.: 2003,
\newblock {\em \aap} {\bf 399}, 133

\bibitem[\protect\astroncite{{Rebassa-Mansergas} et~al.}{2013}]{rebassa13}
{Rebassa-Mansergas}, A., {Schreiber}, M.~R., and {G{\"a}nsicke}, B.~T.: 2013,
\newblock {\em \mnras} {\bf 429}, 3570

\bibitem[\protect\astroncite{{Rebolo} et~al.}{1996}]{rebolo96}
{Rebolo}, R., {Martin}, E.~L., {Basri}, G., {Marcy}, G.~W., and
  {Zapatero-Osorio}, M.~R.: 1996,
\newblock {\em \apjl} {\bf 469}, L53

\bibitem[\protect\astroncite{{Rebull} et~al.}{2006}]{rebull06}
{Rebull}, L.~M., {Stauffer}, J.~R., {Megeath}, S.~T., {Hora}, J.~L., and
  {Hartmann}, L.: 2006,
\newblock {\em \apj} {\bf 646}, 297

\bibitem[\protect\astroncite{{Reiners}}{2007}]{reiners07}
{Reiners}, A.: 2007,
\newblock {\em Astronomische Nachrichten} {\bf 328}, 1034

\bibitem[\protect\astroncite{{Reiners} and {Mohanty}}{2012}]{reiners12}
{Reiners}, A. and {Mohanty}, S.: 2012,
\newblock {\em \apj} {\bf 746}, 43

\bibitem[\protect\astroncite{{Reinhold} et~al.}{2013}]{reinhold13}
{Reinhold}, T., {Reiners}, A., and {Basri}, G.: 2013,
\newblock {\em \aap} {\bf 560}, A4

\bibitem[\protect\astroncite{{Rhode} et~al.}{2001}]{rhode01}
{Rhode}, K.~L., {Herbst}, W., and {Mathieu}, R.~D.: 2001,
\newblock {\em \aj} {\bf 122}, 3258

\bibitem[\protect\astroncite{{Ribas} et~al.}{2005}]{ribas05}
{Ribas}, I., {Guinan}, E.~F., {G{\"u}del}, M., and {Audard}, M.: 2005,
\newblock {\em \apj} {\bf 622}, 680

\bibitem[\protect\astroncite{{Richer} and {Michaud}}{1993}]{richer93}
{Richer}, J. and {Michaud}, G.: 1993,
\newblock {\em \apj} {\bf 416}, 312

\bibitem[\protect\astroncite{{Robrade} et~al.}{2012}]{robrade12}
{Robrade}, J., {Schmitt}, J.~H.~M.~M., and {Favata}, F.: 2012,
\newblock {\em \aap} {\bf 543}, A84

\bibitem[\protect\astroncite{{Ryan} et~al.}{2001}]{ryan01}
{Ryan}, S.~G., {Kajino}, T., {Beers}, T.~C., {Suzuki}, T.~K., {Romano}, D.,
  {Matteucci}, F., and {Rosolankova}, K.: 2001,
\newblock {\em \apj} {\bf 549}, 55

\bibitem[\protect\astroncite{{Sbordone} et~al.}{2010}]{sbordone10}
{Sbordone}, L., {Bonifacio}, P., {Caffau}, E., {Ludwig}, H.-G., {Behara},
  N.~T., {Gonz{\'a}lez Hern{\'a}ndez}, J.~I., {Steffen}, M., {Cayrel}, R.,
  {Freytag}, B., {van't Veer}, C., {Molaro}, P., {Plez}, B., {Sivarani}, T.,
  {Spite}, M., {Spite}, F., {Beers}, T.~C., {Christlieb}, N., {Fran{\c c}ois},
  P., and {Hill}, V.: 2010,
\newblock {\em \aap} {\bf 522}, A26

\bibitem[\protect\astroncite{{Sergison} et~al.}{2013}]{sergison13}
{Sergison}, D.~J., {Mayne}, N.~J., {Naylor}, T., {Jeffries}, R.~D., and {Bell},
  C.~P.~M.: 2013,
\newblock {\em \mnras} {\bf 434}, 966

\bibitem[\protect\astroncite{{Sestito} and {Randich}}{2005}]{sestito05}
{Sestito}, P. and {Randich}, S.: 2005,
\newblock {\em \aap} {\bf 442}, 615

\bibitem[\protect\astroncite{{Shkolnik} et~al.}{2009}]{shkolnik09}
{Shkolnik}, E., {Liu}, M.~C., and {Reid}, I.~N.: 2009,
\newblock {\em \apj} {\bf 699}, 649

\bibitem[\protect\astroncite{{Siess} et~al.}{2000}]{siess2000}
{Siess}, L., {Dufour}, E., and {Forestini}, M.: 2000,
\newblock {\em \aap} {\bf 358}, 593

\bibitem[\protect\astroncite{{Sills} et~al.}{2000}]{sills00}
{Sills}, A., {Pinsonneault}, M.~H., and {Terndrup}, D.~M.: 2000,
\newblock {\em \apj} {\bf 534}, 335

\bibitem[\protect\astroncite{{Silvestri} et~al.}{2006}]{silvestri06}
{Silvestri}, N.~M., {Hawley}, S.~L., {West}, A.~A., {Szkody}, P., {Bochanski},
  J.~J., {Eisenstein}, D.~J., {McGehee}, P., {Schmidt}, G.~D., {Smith}, J.~A.,
  {Wolfe}, M.~A., {Harris}, H.~C., {Kleinman}, S.~J., {Liebert}, J., {Nitta},
  A., {Barentine}, J.~C., {Brewington}, H.~J., {Brinkmann}, J., {Harvanek}, M.,
  {Krzesi{\'n}ski}, J., {Long}, D., {Neilsen}, Jr., E.~H., {Schneider}, D.~P.,
  and {Snedden}, S.~A.: 2006,
\newblock {\em \aj} {\bf 131}, 1674

\bibitem[\protect\astroncite{{Simon} and {Patten}}{1998}]{simon98}
{Simon}, T. and {Patten}, B.~M.: 1998,
\newblock {\em \pasp} {\bf 110}, 283

\bibitem[\protect\astroncite{{Skumanich}}{1972}]{skumanich72}
{Skumanich}, A.: 1972,
\newblock {\em \apj} {\bf 171}, 565

\bibitem[\protect\astroncite{{Soderblom}}{2010}]{soderblom10}
{Soderblom}, D.~R.: 2010,
\newblock {\em \araa} {\bf 48}, 581

\bibitem[\protect\astroncite{{Soderblom} et~al.}{1991}]{soderblom91}
{Soderblom}, D.~R., {Duncan}, D.~K., and {Johnson}, D.~R.~H.: 1991,
\newblock {\em \apj} {\bf 375}, 722

\bibitem[\protect\astroncite{{Soderblom} et~al.}{2013}]{soderblom13}
{Soderblom}, D.~R., {Hillenbrand}, L.~A., {Jeffries}, R.~D., {Mamajek}, E.~E.,
  and {Naylor}, T.: 2013,
\newblock {\em ArXiv e-prints 1311.7024}

\bibitem[\protect\astroncite{{Soderblom} et~al.}{1993}]{soderblom93a}
{Soderblom}, D.~R., {Jones}, B.~F., {Balachandran}, S., {Stauffer}, J.~R.,
  {Duncan}, D.~K., {Fedele}, S.~B., and {Hudon}, J.~D.: 1993,
\newblock {\em \aj} {\bf 106}, 1059

\bibitem[\protect\astroncite{{Soderblom} et~al.}{1995}]{soderblom95}
{Soderblom}, D.~R., {Jones}, B.~F., {Stauffer}, J.~R., and {Chaboyer}, B.:
  1995,
\newblock {\em \aj} {\bf 110}, 729

\bibitem[\protect\astroncite{{Soderblom} et~al.}{1999}]{soderblom99}
{Soderblom}, D.~R., {King}, J.~R., {Siess}, L., {Jones}, B.~F., and {Fischer},
  D.: 1999,
\newblock {\em \aj} {\bf 118}, 1301

\bibitem[\protect\astroncite{{Soderblom} et~al.}{1990}]{soderblom90}
{Soderblom}, D.~R., {Oey}, M.~S., {Johnson}, D.~R.~H., and {Stone}, R.~P.~S.:
  1990,
\newblock {\em \aj} {\bf 99}, 595

\bibitem[\protect\astroncite{{Somers} and {Pinsonneault}}{2014}]{somers14}
{Somers}, G. and {Pinsonneault}, M.: 2014,
\newblock {\em ArXiv e-prints 1402.6333}

\bibitem[\protect\astroncite{{Spada} et~al.}{2011}]{spada11}
{Spada}, F., {Lanzafame}, A.~C., {Lanza}, A.~F., {Messina}, S., and {Collier
  Cameron}, A.: 2011,
\newblock {\em \mnras} {\bf 416}, 447

\bibitem[\protect\astroncite{{Spite} and {Spite}}{1982}]{spite82}
{Spite}, F. and {Spite}, M.: 1982,
\newblock {\em \aap} {\bf 115}, 357

\bibitem[\protect\astroncite{{Stauffer} et~al.}{1999}]{stauffer99}
{Stauffer}, J.~R., {Barrado y Navascu{\'e}s}, D., {Bouvier}, J., {Morrison},
  H.~L., {Harding}, P., {Luhman}, K.~L., {Stanke}, T., {McCaughrean}, M.,
  {Terndrup}, D.~M., {Allen}, L., and {Assouad}, P.: 1999,
\newblock {\em \apj} {\bf 527}, 219

\bibitem[\protect\astroncite{{Stauffer} et~al.}{1998}]{stauffer98}
{Stauffer}, J.~R., {Schultz}, G., and {Kirkpatrick}, J.~D.: 1998,
\newblock {\em \apjl} {\bf 499}, L199

\bibitem[\protect\astroncite{{Steinhauer} and
  {Deliyannis}}{2004}]{steinhauer04}
{Steinhauer}, A. and {Deliyannis}, C.~P.: 2004,
\newblock {\em \apjl} {\bf 614}, L65

\bibitem[\protect\astroncite{{Stelzer} et~al.}{2013}]{stelzer13}
{Stelzer}, B., {Marino}, A., {Micela}, G., {L{\'o}pez-Santiago}, J., and
  {Liefke}, C.: 2013,
\newblock {\em \mnras} {\bf 431}, 2063

\bibitem[\protect\astroncite{{Strong} and {Saba}}{2009}]{strong09}
{Strong}, K.~T. and {Saba}, J.~L.~R.: 2009,
\newblock {\em Advances in Space Research} {\bf 43}, 756

\bibitem[\protect\astroncite{{Takeda} and {Kawanomoto}}{2005}]{takeda05}
{Takeda}, Y. and {Kawanomoto}, S.: 2005,
\newblock {\em PASJ} {\bf 57}, 45

\bibitem[\protect\astroncite{{Telleschi} et~al.}{2005}]{telleschi05}
{Telleschi}, A., {G{\"u}del}, M., {Briggs}, K., {Audard}, M., {Ness}, J.-U.,
  and {Skinner}, S.~L.: 2005,
\newblock {\em \apj} {\bf 622}, 653

\bibitem[\protect\astroncite{{Tognelli} et~al.}{2012}]{tognelli12}
{Tognelli}, E., {Degl'Innocenti}, S., and {Prada Moroni}, P.~G.: 2012,
\newblock {\em \aap} {\bf 548}, A41

\bibitem[\protect\astroncite{{Vauclair}}{1988}]{vauclair88}
{Vauclair}, S.: 1988,
\newblock {\em \apj} {\bf 335}, 971

\bibitem[\protect\astroncite{{Vican}}{2012}]{vican12}
{Vican}, L.: 2012,
\newblock {\em \aj} {\bf 143}, 135

\bibitem[\protect\astroncite{{Vilhu} and {Walter}}{1987}]{vilhu87}
{Vilhu}, O. and {Walter}, F.~M.: 1987,
\newblock {\em \apj} {\bf 321}, 958

\bibitem[\protect\astroncite{{Wagoner} et~al.}{1967}]{wagoner67}
{Wagoner}, R.~V., {Fowler}, W.~A., and {Hoyle}, F.: 1967,
\newblock {\em \apj} {\bf 148}, 3

\bibitem[\protect\astroncite{{Walkowicz} and {Basri}}{2013}]{walkowicz13}
{Walkowicz}, L.~M. and {Basri}, G.~S.: 2013,
\newblock {\em \mnras} {\bf 436}, 1883

\bibitem[\protect\astroncite{{Wallerstein} et~al.}{1965}]{wallerstein65}
{Wallerstein}, G., {Herbig}, G.~H., and {Conti}, P.~S.: 1965,
\newblock {\em \apj} {\bf 141}, 610

\bibitem[\protect\astroncite{{Wolk} et~al.}{2005}]{wolk05}
{Wolk}, S.~J., {Harnden}, Jr., F.~R., {Flaccomio}, E., {Micela}, G., {Favata},
  F., {Shang}, H., and {Feigelson}, E.~D.: 2005,
\newblock {\em \apjs} {\bf 160}, 423

\bibitem[\protect\astroncite{{Wright} et~al.}{2011}]{wright11}
{Wright}, N.~J., {Drake}, J.~J., {Mamajek}, E.~E., and {Henry}, G.~W.: 2011,
\newblock {\em \apj} {\bf 743}, 48

\bibitem[\protect\astroncite{{Zanni} and {Ferreira}}{2013}]{zanni13}
{Zanni}, C. and {Ferreira}, J.: 2013,
\newblock {\em \aap} {\bf 550}, A99

\bibitem[\protect\astroncite{{Zuckerman} and {Webb}}{2000}]{zuckerman00}
{Zuckerman}, B. and {Webb}, R.~A.: 2000,
\newblock {\em \apj} {\bf 535}, 959

\end{thebibliography}

%%-----------------------------
%%      your bibliography
%%-----------------------------
%\begin{thebibliography}{99}
%\bibitem[1994]{alref1} Aalto, S. \etal  1994, A\&A, 286, 365.
%%% Using \cite{Bei} in the text
%\bibitem[1986]{Bei} Beichman, C.A., Neugebauer, G., Habing,
%   H., Clegg, P.E. \& Chester, T.C. 1988, editors, {\it ``IRAS Catalogs and
%   Atlases: Explanatory Supplement''}, NASA RP-1190 (Washington: NASA)
%\bibitem[1987]{ref1987} Beichman, C.A. 1987, ARA\&A, 25, 521
%\bibitem[1987]{so1987} Soifer, B.T., Houck, J.R. and Neugebauer, G. 1987, ARAA, 25, 187
%\end{thebibliography}
\end{document}